\documentclass[fleqn,usenatbib]{mnras}

\usepackage{newtxtext,newtxmath}
\usepackage{mdframed}

\usepackage[T1]{fontenc}
\usepackage{ae,aecompl}
\usepackage{comment}

\usepackage{graphicx}	
\usepackage[dvipsnames]{xcolor}
\usepackage{amsmath}	
\usepackage{physics}
\usepackage{siunitx}
\usepackage{xspace}
\usepackage{tikz}
\usepackage[normalem]{ulem}
\usepackage{multirow}
\usepackage{booktabs}
\usepackage{bbold}

\usepackage[capitalize]{cleveref}
\Crefname{equation}{Equation}{Equations}
\crefname{equation}{Eq.}{Eqs.}
\Crefname{figure}{Figure}{Figures}
\crefname{figure}{Fig.}{Figs.}
\Crefname{table}{Table}{Tables}
\crefname{table}{Tab.}{Tabs.}
\Crefname{section}{Section}{Sections}
\crefname{section}{Sec.}{Secs.}

\def\P{\mathcal{P}}

\def\3x2pt{3x2pt}

\newcommand{\xip}{\xi_+}
\newcommand{\xim}{\xi_-}
\newcommand{\xipm}{\xi_\pm}
\newcommand{\gammat}{\gamma_{\rm t}}
\newcommand{\wtheta}{w(\theta)}
\newcommand{\vs}{\textit{vs}\xspace}
\newcommand{\lcdm}{{$\Lambda$CDM}\xspace}

\title[DES internal consistency with PPD]{Dark Energy Survey internal consistency tests of the joint cosmological probes analysis with posterior predictive distributions}

\author[DES Collaboration]{
\parbox{\textwidth}{
\Large
C.~Doux,$^{1}\star$
E.~Baxter,$^{2}\dag$
P.~Lemos,$^{3}$
C.~Chang,$^{4,5}$
A.~Alarcon,$^{6}$
A.~Amon,$^{7}$
A.~Campos,$^{8}$
A.~Choi,$^{9}$
M.~Gatti,$^{10}$
D.~Gruen,$^{11,7,12}$
M.~Jarvis,$^{1}$
N.~MacCrann,$^{13}$
Y.~Park,$^{14}$
J.~Prat,$^{4}$
M.~M.~Rau,$^{8}$
M.~Raveri,$^{5}$
S.~Samuroff,$^{8}$
J.~DeRose,$^{15,16}$
W.~G.~Hartley,$^{17}$
B.~Hoyle,$^{18,19,20}$
M.~A.~Troxel,$^{21}$
J.~Zuntz,$^{22}$
T.~M.~C.~Abbott,$^{23}$
M.~Aguena,$^{24,25}$
S.~Allam,$^{26}$
J.~Annis,$^{26}$
S.~Avila,$^{27}$
D.~Bacon,$^{28}$
E.~Bertin,$^{29,30}$
S.~Bhargava,$^{31}$
D.~Brooks,$^{3}$
D.~L.~Burke,$^{7,12}$
M.~Carrasco~Kind,$^{32,33}$
J.~Carretero,$^{10}$
R.~Cawthon,$^{34}$
M.~Costanzi,$^{35,36}$
L.~N.~da Costa,$^{25,37}$
M.~E.~S.~Pereira,$^{38}$
S.~Desai,$^{39}$
H.~T.~Diehl,$^{26}$
J.~P.~Dietrich,$^{18}$
P.~Doel,$^{3}$
S.~Everett,$^{16}$
I.~Ferrero,$^{40}$
P.~Fosalba,$^{41,42}$
J.~Frieman,$^{26,5}$
J.~Garc\'ia-Bellido,$^{27}$
D.~W.~Gerdes,$^{43,38}$
T.~Giannantonio,$^{13,44}$
R.~A.~Gruendl,$^{32,33}$
J.~Gschwend,$^{25,37}$
G.~Gutierrez,$^{26}$
S.~R.~Hinton,$^{45}$
D.~L.~Hollowood,$^{16}$
K.~Honscheid,$^{9,46}$
E.~M.~Huff,$^{47}$
D.~Huterer,$^{38}$
B.~Jain,$^{1}$
D.~J.~James,$^{48}$
E.~Krause,$^{49}$
K.~Kuehn,$^{50,51}$
N.~Kuropatkin,$^{26}$
O.~Lahav,$^{3}$
C.~Lidman,$^{52,53}$
M.~Lima,$^{24,25}$
M.~A.~G.~Maia,$^{25,37}$
F.~Menanteau,$^{32,33}$
R.~Miquel,$^{54,10}$
R.~Morgan,$^{34}$
J.~Muir,$^{7}$
R.~L.~C.~Ogando,$^{25,37}$
A.~Palmese,$^{26,5}$
F.~Paz-Chinch\'{o}n,$^{13,33}$
A.~A.~Plazas,$^{55}$
E.~Sanchez,$^{56}$
V.~Scarpine,$^{26}$
M.~Schubnell,$^{38}$
S.~Serrano,$^{41,42}$
I.~Sevilla-Noarbe,$^{56}$
M.~Smith,$^{57}$
E.~Suchyta,$^{58}$
M.~E.~C.~Swanson,$^{33}$
G.~Tarle,$^{38}$
C.~To,$^{11,7,12}$
D.~L.~Tucker,$^{26}$
T.~N.~Varga,$^{19,20}$
J.~Weller,$^{19,20}$
and R.D.~Wilkinson$^{31}$
\begin{center} (DES Collaboration) \end{center}
}
}

\usepackage{eso-pic}

\AddToShipoutPictureBG*{%
  \AtPageUpperLeft{%
    \hspace{0.75\paperwidth}%
    \raisebox{-3.5\baselineskip}{%
      \makebox[0pt][l]{\textnormal{DES-2020-0553}}
}}}%

\AddToShipoutPictureBG*{%
  \AtPageUpperLeft{%
    \hspace{0.75\paperwidth}%
    \raisebox{-4.5\baselineskip}{%
      \makebox[0pt][l]{\textnormal{FERMILAB-PUB-20-547-AE}}
}}}%

\date{Accepted XXX. Received YYY; in original form ZZZ}

\pubyear{2020}

\begin{document}
\label{firstpage}
\pagerange{\pageref{firstpage}--\pageref{lastpage}}

\maketitle

\begin{abstract}
Beyond-\lcdm physics or systematic errors may cause subsets of a cosmological data set to appear inconsistent when analyzed assuming \lcdm.  We present an application of internal consistency tests to measurements from the Dark Energy Survey Year 1 (DES~Y1) joint probes analysis.  Our analysis relies on computing the posterior predictive distribution (PPD) for these data under the assumption of \lcdm.  We find that the DES~Y1 data have an acceptable goodness of fit to \lcdm, with a probability of finding a worse fit by random chance of ${p = 0.046}$.  Using numerical PPD tests, supplemented by graphical checks, we show that most of the data vector appears completely consistent with expectations, although we observe a small tension between large- and small-scale measurements.  A small part (roughly 1.5\%) of the data vector shows an unusually large departure from expectations; excluding this part of the data has negligible impact on cosmological constraints, but does significantly improve the $p$-value to 0.10. The methodology developed here will be applied to test the consistency of DES Year 3 joint probes data sets.
\end{abstract}


\makeatletter
\def \blfootnote{\xdef\@thefnmark{}\@footnotetext}
\makeatother

\blfootnote{$^{\star}$ E-mail: cdoux@sas.upenn.edu}
\blfootnote{$^{\dag}$ E-mail: ericjbax@gmail.com}

\section{Introduction}

Several recent cosmological measurements appear to be in mild to severe tension in the context of the standard cosmological constant and cold dark matter (\lcdm) model.  For instance, the value of $H_0$ inferred from the cosmic microwave background \citep[CMB,][]{2020A&A...641A...6P} and from the cosmic distance ladder \citep{Riess:2019} are discrepant at roughly the $5\sigma$ level  \citep[e.g.][]{Bernal:2016,Feeney:2018, Aylor:2019}.  Similarly, the value of $\sigma_8$ inferred from the cosmic microwave background and from large-scale structure, in particular weak lensing measurements, are discrepant at roughly the $3\sigma$ level  \citep[e.g][]{Battye:2015,MacCrann:2015,Raveri:2016,Hildebrandt:2016,3x2pt,Raveri:2019,2020A&A...634A.127A,2020A&A...638L...1J,2020arXiv200715632H,2020MNRAS.499.4638P}.
These tensions could be indicative either of a breakdown in the standard cosmological model, or of systematics impacting various analyses.  Given these possibilities, identifying and quantifying cosmological tensions is of prime importance.  We make a somewhat artificial distinction between {\it external} tensions---those between different experiments---and {\it internal} tensions---those between the measurements of a single experiment.  In practice, correlations between the data are common in the case of internal tensions, but are rarer for external tensions.

In this work, we explore internal tensions in the Dark Energy Survey \citep[DES;][]{DES} Year 1 (Y1) measurements of two-point functions of large-scale structure \citep{3x2pt}.  The Dark Energy Survey is a six-year optical imaging survey of 5000 square degrees of the southern sky.  The analysis of \citet{3x2pt} derived measurements of galaxy positions and galaxy shear from first year observations of DES, and used these to measure three two-point functions: the auto-correlations of galaxy shear and of galaxy positions, and the cross-correlation of galaxy position with galaxy shear.  Cosmological constraints were then obtained by fitting this so-called \3x2pt combination of correlation functions.

There are a few reasons why tests of internal consistency of the DES data are important and timely.  First, as mentioned previously, measurements of $\sigma_8$ from surveys of large-scale structure tend to be lower than the value inferred from primary anisotropies in the cosmic microwave background.  If this tension is due to a true breakdown in \lcdm---such as a departure from the expected growth of structure---then DES data alone might be expected to be internally inconsistent assuming \lcdm.  Second, all weak lensing surveys necessarily suffer from systematic errors \citep[see, e.g.,][]{Chang:2019}, thus introducing uncertainties which must be accounted for.  Systematic errors in the data, such as unaccounted photometric redshift biases, are likely to manifest as internal inconsistency.  Finally, one of the aims of this analysis is to develop the methodology and specific data tests that will be applied to the forthcoming analysis of Year 3 (Y3) data from DES.

We address the question of whether the DES Y1 \3x2pt measurements are self-consistent using posterior predictive methods.
The posterior predictive distribution \citep[PPD, see, e.g.,][for a review]{Gelman:2013} is the distribution of possible new data, conditioned on observed data, given an underlying model. By comparing the PPD to the observed data, we can assess the degree to which the observed data are internally consistent in the context of \lcdm.
Several recent works have adopted the PPD as a means to examine consistency of cosmological datasets \cite[e.g.][]{Feeney:2019,5x2pt}.
\cite{Kohlinger:2018sxx} proposed a test of internal consistency based on three different tests, that occur at various levels of the analysis: a global test, based on ratios of Bayesian evidences \citep{Marshall:2006}, a parameter difference test that occurs in parameter space, and a PPD test on data space. Later, \cite{Handley:2019wlz} showed that the test based on the evidence ratio is proportional to the prior volume, and substituted it with a test based on the \textit{suspiciousness} statistic \citep[extended to correlated data sets in][]{Lemos:2019txn}.
Here, we focus on identifying potential subsets of the DES Y1 data in tension with each other, for which the PPD tests are particularly well-suited since they operate entirely in data space.

Our approach is to split the DES Y1 \3x2pt data into subsets motivated by considerations of possible systematics, as well as by considerations of possible extensions to \lcdm. We first evaluate the goodness of fit of these subsets of data to \lcdm using the standard PPD formalism. Next, we use the PPD to perform \textit{consistency} tests where we evaluate the goodness of fit of some subset of the data {\it conditioned} on the observed data from another, disjoint subset.  For instance, we consider the likelihood of the measured two-point functions at large scales, conditioned on their observed values at small scales.  This test in effect determines whether the large and small scale measurements are consistent.  

The paper is organized as follows.  In \cref{sec:des_data} we describe the DES Y1 \3x2pt measurements; in \cref{sec:ppd} we give an overview of the PPD framework and application to DES Y1 data; in \cref{sec:application} we present the results of the application of this framework to the DES Y1 measurements; in \cref{sec:y3} we lay out our plan for the upcoming DES Y3 analysis; we conclude in \cref{sec:discussion}.

\section{The DES Y1 3x2pt measurements}
\label{sec:des_data}

In this section we briefly describe the \3x2pt data vector from the DES~Y1 analysis; more details can be found in \citet{3x2pt}.  From the DES imaging data, galaxy positions and shears are measured.  The galaxy samples are divided into two samples: "lenses" and "sources."  The lens galaxies are selected using the \texttt{redMaGiC} algorithm \citep*{redmagic}, and have tightly constrained redshifts.  The source galaxy sample, which extends to higher redshift than the lenses, have shapes measured using \texttt{METACALIBRATION} \citep{metacal} and \texttt{Im3SHAPE}, as described in \citet*{Zuntz:2018}.
Using the galaxy position measurements and the galaxy shear measurements (for the sources only), three two point correlations functions were computed: shear-shear, position-shear and position-position.
Each two-point correlation was computed as a function of angular separation, $\theta$. Since \emph{cosmic shear} is a spin-2 field, the shear-shear correlation was divided into two components, $\xip(\theta)$, and $\xim(\theta)$.  For notational convenience, we refer to the position-position correlation, or \emph{clustering}, as $\wtheta$, and to the galaxy-shear correlation, also referred to as \emph{galaxy-galaxy lensing}, as $\gammat(\theta)$.  The lens and source samples were divided into five and four tomographic redshift bins, respectively, and the auto and cross-correlations between the bins were measured. While all correlation functions were computed at twenty fixed (logarithmic) angular bins between 2.5 and 250 arcminutes, measurements at small scales were not included in the analysis due to the presence of effects that could make the DES Y1 model an inaccurate description of the data in the small-scale regime (such as nonlinear galaxy bias, baryonic effects on the matter power spectrum, etc).  Details of the measurements of these correlation functions can be found  in \citet{ElvinPoole:2018ew,Troxel:2018gr}; \citet*{Prat:2018im}.  The full \3x2pt data vector includes all two-point measurements, in all redshift bin combinations, across a range of angular scales.

\section{Posterior Predictive Distribution}
\label{sec:ppd}

In this section, we present an overview of the PPD formalism (\cref{sec:ppd_overview}), discuss the choice of test statistic (\cref{sec:low_pvalues}), its application to DES Y1 data for \textit{goodness-of-fit} and \textit{consistency} tests (\cref{sec:PPD_on_DESY1}) and our sampling strategy (\cref{sec:sampling}).
In particular, we identify a potential caveat associated with the standard choice of $\chi^2$ statistic when testing the consistency of two experiments whose posteriors have little overlap. We illustrate this problem with a toy model in \cref{sec:low_pvalues} and present a solution applicable to DES Y1 data in \cref{sec:PPD_on_DESY1}.

\subsection{Overview}
\label{sec:ppd_overview}

The posterior predictive distribution (PPD) is the distribution of possible (unobserved) data realizations from an experiment, given the posterior on parameters, $\Theta$, of the model, $M$, obtained from the observed data. For observed data, $d_{\rm obs}$, the model posterior is $P(\Theta | d_{\rm obs}, I)$, where $I$ represents all prior information, such as the form of the likelihood and priors.  The posterior predictive distribution function for unobserved data $d_{\rm rep}$ is then\footnote{We follow the notation of \citet{Gelman:2013} in referring to the \textit{replicated}, unobserved PPD repetitions of the data as $d_{\rm rep}$.} 
\begin{equation}
P(d_{\rm rep} | d_{\rm obs}, I) = \int \dd\Theta P(d_{\rm rep} | d_{\rm obs}, \Theta, I) P(\Theta | d_{\rm obs}, I).
\label{eq:ppd_integral}
\end{equation}
In the case that $d_{\rm rep}$ and $d_{\rm obs}$ are conditionally independent given $\Theta$, then $P(d_{\rm rep} | d_{\rm obs}, \Theta, I) = P(d_{\rm rep} | \Theta, I)$, i.e. the data likelihood.  We will consider both this case and the case where $d_{\rm rep}$ and $d_{\rm obs}$ are not conditionally independent below. Hereafter, we drop $I$ for conciseness.

The analytic computation of the the integral in \cref{eq:ppd_integral} is cumbersome. Furthermore, the quantity on the left is a probability density, and therefore it requires to be normalized to provide a meaningful statement about the statistical significance of the internal tensions in a data set. One straightforward possibility is to take the ratio ${R = { P(d_{\rm rep} | d_{\rm obs}) / P(d_{\rm rep})}}$,
where we can identify the denominator as the Bayesian evidence for $d_{\rm rep}$. However, as pointed out by \cite{Lemos:2019txn}, this is equivalent to the Bayes' ratio \citep{Marshall:2006}, which is not suitable for the case of wide and uninformative priors. 
In practice, rather than compute the integral in \cref{eq:ppd_integral}, one typically draws realizations of $d_{\rm rep}$ from $P(d_{\rm rep} | d_{\rm obs}, \Theta)$ at each point $\Theta$ in a Markov chain sampling the posterior $P(\Theta | d_{\rm obs}, I)$.  These realizations of $d_{\rm rep}$, i.e. samples from the PPD, can then be compared directly to the observed data $d_{\rm obs}$ to look for signs of discrepancy.

Below, we will explore graphical and numerical approaches to comparing $d_{\rm rep}$ and $d_{\rm obs}$. Indeed, a powerful application of the PPD framework is to simply plot the observed PPD realization on top of the actual data, as done in \cref{fig:ppd_3x2} for instance.  If the true data look significantly different from the PPD realizations, that could be a sign that the data are inconsistent with the assumed model. However, this approach does not allow one to quantify consistency and is mainly used here to provide insight into useful splits of the full data vector. A common way to quantify the level of tension between $d_{\rm rep}$ and $d_{\rm obs}$ is to use a test statistic, $T(d,\Theta)$, that can be computed for both $d_{\rm rep}$ and $d_{\rm obs}$, and which may be a function of the parameters~$\Theta$.  A $p$-value can then be associated with the comparison between $d_{\rm rep}$ and $d_{\rm obs}$ via
\begin{equation}
p = P(T(d_{\rm rep},\Theta) > T(d_{\rm obs}, \Theta)| d_{\rm obs}).
\label{eq:pvalue}
\end{equation}
In other words, $p$ is the probability of getting a higher test statistic for PPD realizations than $T(d_{\rm obs}, \Theta)$ by random chance.  A very low $p$-value (say less that 0.01) would then be indicative that $d_{\rm obs}$ was unlikely, while a high $p$-value (say greater than 0.99) could be indicative of a problem in the model, such as an overestimate of the noise covariance.  Following \citet{3x2pt}, we will adopt a $p$-value threshold of $p=0.01$: if $p > 0.01$, we will consider the data to be in reasonable consistency.
In practice, the $p$-value can be computed easily from a Markov chain sampling the posterior $P(\Theta | d_{\rm obs})$.  At each set of parameters $\Theta_i$ in the chain, one draws a realization $d_{{\rm rep},i}$ of $d_{\rm rep}$ using the (possibly conditional) likelihood $P(d_{\rm rep} | d_{\rm obs}, \Theta_i)$.  Values of $T(d_{\rm obs},\Theta_i)$ and $T(d_{{\rm rep},i},\Theta_i)$ are then computed.  The $p$-value is simply the fraction of samples $\Theta_i$ for which $T(d_{{\rm rep},i},\Theta_i) > T(d_{\rm obs}, \Theta_i)$. 
One of the challenges of a PPD analysis is selecting an appropriate test statistic, $T(d,\Theta)$, as seen in the next section.

Compared to other tension metrics, the PPD has several advantages.  First, unlike evidence ratio-based tests \citep[e.g.][]{Marshall:2006}, the PPD does not require specifying an alternative model.  This is often desirable, since the choice of alternate model is not always clear, especially in the case of cosmological analyses.  Second, the comparison between data and PPD realizations occurs entirely in data space rather than in parameter space.  This means that the PPD is particularly well suited to identifying particular parts of the data that may be unusually discrepant with the model.  Finally, if the posterior is likelihood-dominated (as opposed to prior-dominated), the PPD realizations will not depend on the prior volume.  This is not the case for e.g. evidence ratios, for which the choice of prior outside the likelihood-dominated region is important.

\subsection{Choice of test statistic for computing $p$-value}
\label{sec:low_pvalues}

As discussed above, assigning a $p$-value to the results of a PPD test requires a choice of test statistic.  For high-dimensional data---in particular Gaussian data---a common choice is $\chi^2$, i.e.
\begin{equation}
    T(d,\Theta) = \qty(d - \mu(\Theta))^\intercal \mathbf{C}^{-1} \qty(d-\mu(\Theta)),
    \label{eq:test_stat}
\end{equation}
where $\mu(\Theta)$ is the model evaluated at parameter values $\Theta$ and $\mathbf{C}$ is the covariance matrix.
The use of $\chi^2$ as the test statistic, however, can bias the $p$-value low when testing the consistency of two experiments---which can be disjoint subsets of a data vector---that constrain very different volumes of the parameter space. In this case, the replicas of one experiment conditioned on the other can naturally yield $p$-values that are not uniformly distributed in $[0,1]$ over (consistent) data realizations, but are skewed towards lower values. Thus these should be interpreted with caution when using them for consistency tests, as we illustrate with a toy model in the following subsection \citep[see also][for a discussion about non-uniform posterior predictive $p$-values]{Gelman:2013jm}. In \cref{sec:PPD_on_DESY1}, we show how PPD tests can be repeated on simulated DES Y1 data to calibrate $p$-values and distinguish this effect from true tensions. We will therefore rely on calibrated $p$-values, denoted $\tilde{p}$, to test internal consistency.

\subsubsection{A toy model example}

\begin{figure*}
    \centering
    \includegraphics[scale = 0.5]{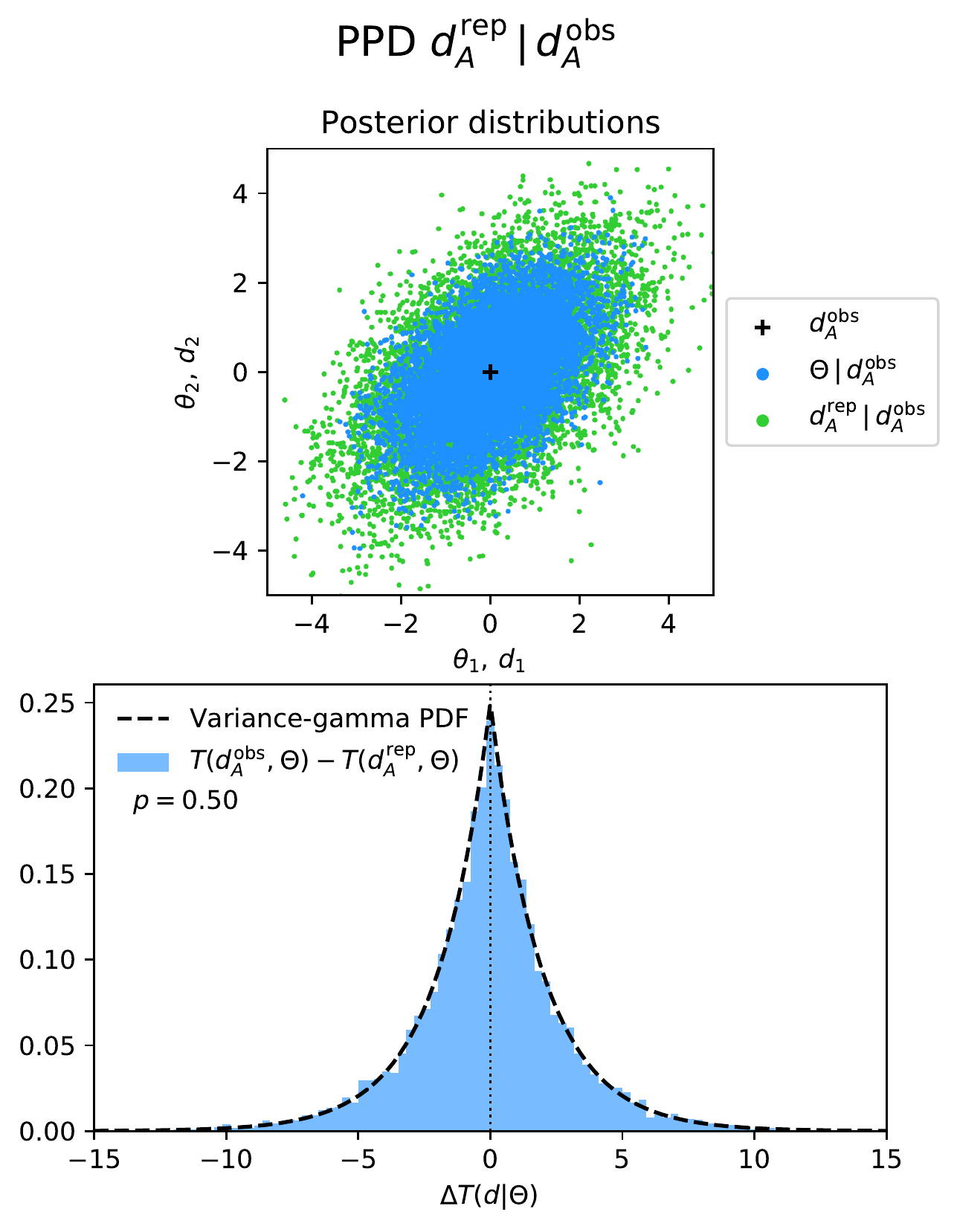}
    \qquad
    \includegraphics[scale = 0.5]{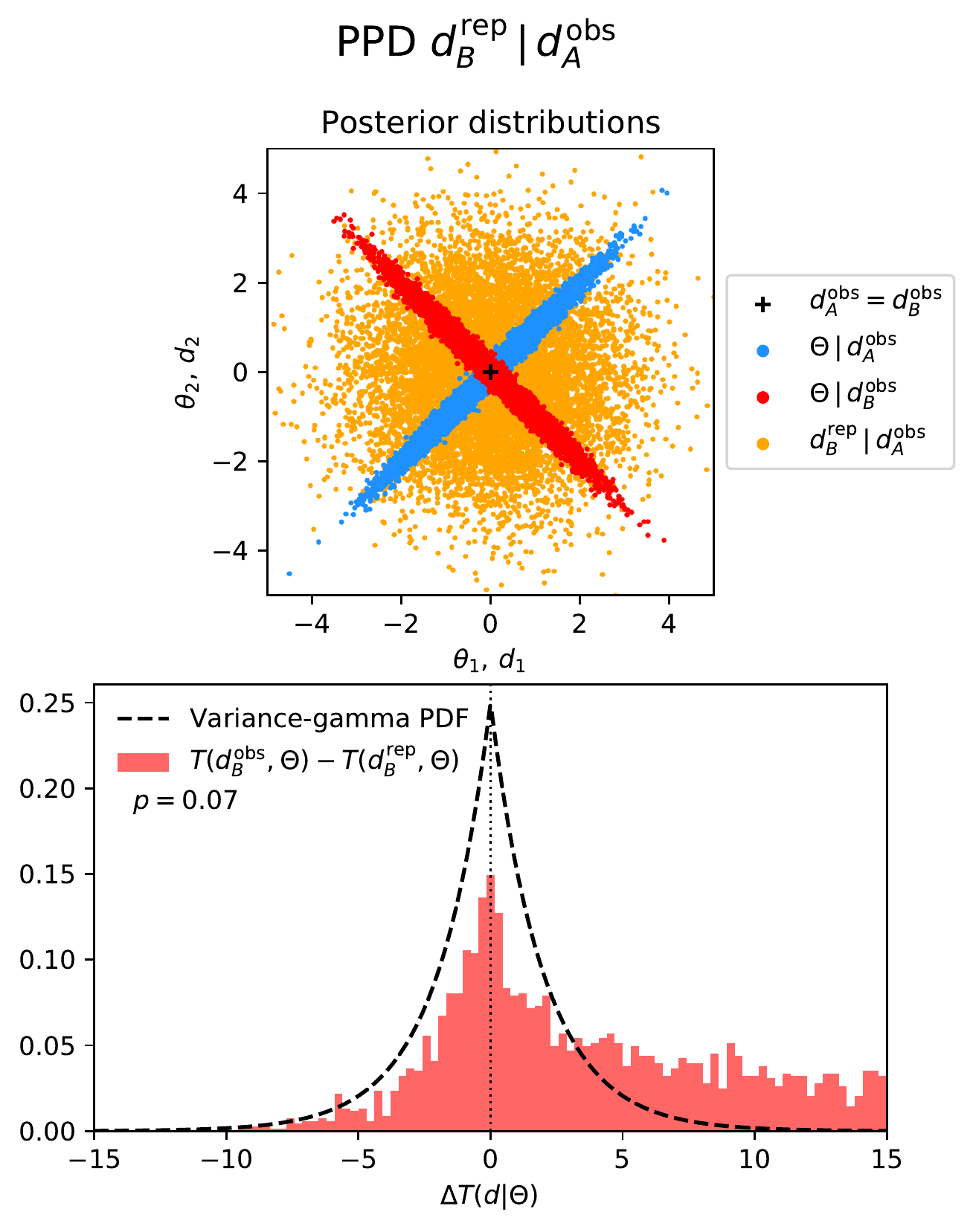}
    \caption{Illustration of the challenges of using $\chi^2$ as a test statistic for calculating a $p$-value from the PPD.  Top panels show draws from the parameter posteriors, while bottom panels show histograms of the difference of computed test statistics \citep[with the probability density function of the difference of independent $\chi^2$ statistics, given by the variance-gamma distribution developed by][]{ 10.2307/2353303}.  In the panels at left, we consider PPD realizations $d_A^{\rm rep}$ from the same experiment used to generate the observed data, i.e. $d_A^{\rm obs}$.  In the panels at right, we consider PPD realizations of a different experiment, $d_B^{\rm rep}$, than the original data, $d_A^{\rm obs}$.  In the latter case, we can make the $p$-value arbitrarily small by reducing the fractional overlap in the posteriors from $A$ and $B$.}
    \label{fig:ppd_chi2_illustration}
\end{figure*}

For purposes of illustration, we consider two independent experiments, $A$ and $B$, that both make two-dimensional measurements, respectively $d_A = (d_1^A, d_2^A)$ and $d_B = (d_1^B, d_2^B)$. We now suppose both experiments to be normally distributed with means $\mu_A$ and $\mu_B$, unit variance per component and covariances $\rho_A$ and $\rho_B$ between components, such that the covariance of $d_A$ is
\begin{equation}
    \mathbf{C}_A = \mqty( 1 & \rho_A \\ \rho_A & 1),
\end{equation}
and similarly for $d_B$.
Problems with using $\chi^2$ as a test statistic emerge when the parameter space is at least two-dimensional.  We therefore consider a two-parameter model for the data, that is specified by parameters $\Theta=(\theta_{1}, \theta_{2})$, with true values ${\Theta_0=(0,0)}$, and such that $\mu_A = \mu_B = \Theta$. In other words, the expectation values of the data points are the parameter values.
Assuming flat priors over parameters, the likelihoods and posteriors are normal distributions we can easily sample from. Finally, we imagine having measurements for each experiment that coincide with their fiducial values, $d_A^{\rm obs}=d_B^{\rm obs}=(0,0)$, such that they are perfectly consistent.

\begin{itemize}
    \item \textit{Goodness-of-fit test.} In the left of \cref{fig:ppd_chi2_illustration}, we perform a \textit{goodness-of-fit} test for experiment $A$ alone using the PPD $P(d_A^{\rm rep}|d_A^{\rm obs})$, which is the distribution of possible future {\it replicated} measurements of $d_A$, given $d_A^{\rm obs}$. We generate a sample of parameters $\Theta$ drawn from the posterior $P(\Theta|d_A^{\rm obs})$ (blue points in the upper left panel) and, at each point $\Theta_i$, we draw a sample from the PPD (green points) by sampling the likelihood ${P(d_A^{\rm rep}|\Theta_i)}$. Here, we assumed $\rho_A=0.5$. Note that the PPD realizations include the uncertainty on the posterior as well as the uncertainty from the data likelihood; this is why the distribution of green points is broader than that of the blue points. In the lower left panel, we show the histogram of the difference of test statistic values from this mock analysis, ${T(d_A^{\rm obs}, \Theta_i)-T(d_A^{\rm rep}, \Theta_i)}$, using the test statistic from \cref{eq:test_stat} with covariance $\mathbf{C}_A$. Comparing the test statistics from $d_A^{\rm obs}$ and from $d_A^{\rm rep}$ yields $p$-value of $p = 0.5$, which is reasonable, given that we know the data are consistent, and the PPD is working as expected. 

    \item \textit{Consistency test.} In the right of \cref{fig:ppd_chi2_illustration}, we perform a \textit{consistency} test between experiments $A$ and $B$ using the PPD $P(d_B^{\rm rep}|d_A^{\rm obs})$. Since experiments $A$ and $B$ are independent, the likelihood is $P(d_B|d_A,\Theta)=P(d_B|\Theta)$, i.e. the normal distribution with mean $\mu_B=\Theta$ and covariance $\mathbf{C}_B$. The problem we highlight here occurs when the overlap of the posteriors from experiments $A$ and $B$ is small. Therefore, we set $\rho_A = 0.99$ and $\rho_B = -0.99$, such that the posteriors of experiments $A$ and $B$, shown respectively in blue and red in the upper right panel, have little overlap around the true parameters.  At each point in the posterior sample for $d_A^{\rm obs}$, we draw realizations from the data likelihood for experiment $B$ in order to generate $d_B^{\rm rep}$ (orange points). This sample can be thought of as combining the posterior uncertainty from experiment $A$ (i.e. the spread of the blue points), with the likelihood uncertainty from experiment $B$ (i.e. the spread of the red points). We see now that, on average, the orange points are far from the red points (as measured with $\chi^2$ using $\mathbf{C}_B$), since the parameter values allowed by the posterior of experiment $A$ are typically far from the posterior of experiment $B$. The histogram of the difference of test statistics, ${T(d_B^{\rm obs}, \Theta_i)-T(d_B^{\rm rep}, \Theta_i)}$, is asymmetric with a tail at higher values, as shown in the lower right panel. As a consequence, the $p$-value for this comparison will be low; we find $p \approx 0.07$.  We can make the $p$-value even lower by increasing $\rho_A$ and decreasing $\rho_B$.
\end{itemize}

\subsubsection{Conclusions from toy model}

We have therefore shown how two experiments that we know are consistent (they are generated from the same true parameter values) can be made to have an arbitrarily low $p$-value if their posteriors do not overlap significantly in parameter space.
We emphasize that this difficulty
emerges because of the choice of test statistic, not because of some fundamental shortcoming of the PPD.  The PPD realizations in orange in the right panel of \cref{fig:ppd_chi2_illustration} are true possible realizations of future measurements of experiment $B$, given our knowledge from experiment $A$. The reason why $p$-values appear to be biased low in the second kind of test is that $T(d^{\rm rep}_B, \Theta_i)$ is likely to be smaller than $T(d^{\rm obs}_B,\Theta_i)$, since $d^{\rm rep}_B$ {\it was generated assuming the true parameter values are $\Theta_i$}. In principle, it may be possible to design some test statistic that does not suffer from this complication; we leave this to future work.

In practice, however, the fact that {\it consistency} tests between two experiments $A$ and $B$ ($d_B^{\rm rep}\vert d_A$)---as opposed to the {\it goodness-of-fit} tests of a single experiment $A$ ($d_A^{\rm rep}\vert d_A$)---tend to bias the $p$-values low---i.e. skew the distribution of $p$-values over data realizations towards low values---means that the PPD tests measure degrees of consistency conservatively. This can be seen as an advantage since we want to be careful about claiming internal consistency. Moreover, most tests we will perform in the following sections do not suffer from this problem as much as our toy model because posteriors are generally not as discrepant.
We will, however, consider a method to calibrate $p$-values in the next section in order to eliminate confusion between this effect and real tensions, and rely on calibrated $\tilde{p}$-values for all tests, in order to facilitate the interpretation of our results.

\subsection{Considerations for application to DES Y1 data}
\label{sec:PPD_on_DESY1}

\subsubsection{DES Y1 likelihood and PPD tests}

The DES \3x2pt analysis, described in \citet*{Krause:2017}, adopts a Gaussian likelihood for the \3x2pt data vector \citep[see, e.g.,][for a discussion of this approximation]{2018MNRAS.473.2355S,2020OJAp....3E..11L}.  The Gaussian likelihood, $\mathcal{L}$, is determined by the expectation value of the data vector at parameters $\Theta$, $\mu(\Theta)$, and by a covariance matrix~$\mathbf{C}$,
\begin{eqnarray}
    \mathcal{L}(d_{\rm obs}|\Theta) & = & \mathcal{N} (\mu(\Theta), \mathbf{C}) \\ & \propto & \exp\qty[ - \frac{1}{2} \qty(d_{\rm obs}-\mu(\Theta))^\intercal \mathbf{C}^{-1} \qty(d_{\rm obs}-\mu(\Theta)) ],
\end{eqnarray}
where $\mathcal{N} (\vb{\mu}, \mathbf{C})$ the probability distribution function of a multivariate Gaussian variable.
We now distinguish two types of tests, depending on whether $d_{\rm rep}$ and $d_{\rm obs}$ refer to the same subset of the data vector (goodness-of-fit test) or different, disjoint subsets (consistency test).
\begin{itemize}
    \item \textit{Goodness-of-fit test.} In this case, $d_{\rm rep}$ is considered to be future, independent realizations of the same observable as $d_{\rm obs}$, in which case $d_{\rm rep}$ does not depend on $d_{\rm obs}$. Therefore, ${P(d_{\rm rep}| d_{\rm obs}, \Theta)=\mathcal{L}(d_{\rm rep}| \Theta)}$ and the PPD $p$-value in \cref{eq:pvalue} can be thought of as a Bayesian goodness-of-fit test, analogous to the classical $\chi^2$ goodness-of-fit test, but including uncertainty over model parameters.
    
    \item \textit{Consistency test.} In this case, $d_{\rm rep}$ and $d_{\rm obs}$ correspond to disjoint, but correlated subsets of the full \3x2pt data vector.  For instance, $d_{\rm obs}$ can consist of cosmic shear and clustering measurements while $d_{\rm rep}$ can refer to galaxy-galaxy lensing observations, i.e. $d_{\rm obs}=\qty{\xipm(\theta),w(\theta)}$ and $d_{\rm rep}=\qty{\gammat(\theta)}$. In this example, the conditional likelihood ${P(d_{\rm rep}| d_{\rm obs}, \Theta)}$ is the distribution of possible realizations of $\gammat(\theta)$ data, given that the measurements of $\xipm(\theta)$ and $w(\theta)$ are known and that the model parameters are known to be $\Theta$. We will, however, consider other ways of splitting the data vector.
    Given our assumption of a multivariate Gaussian likelihood for the data, the distribution $P(d_{\rm rep} | \Theta, d_{\rm obs})$ is also a multivariate Gaussian, the mean and covariance of which can be computed as follows.  For a partition of the full data vector into disjoint subsets $d_1$ and $d_2$, we have
    \begin{equation}
    \mqty(d_1 \\ d_2) \sim \mathcal{N} \qty( \mqty(\mu_1 \\ \mu_2), \mqty(\mathbf{C}_{11} & \mathbf{C}_{12} \\ \mathbf{C}_{21} & \mathbf{C}_{22})),
    \end{equation}
    where $\mu_{i}$ is the mean of $d_{i}$, and $\mathbf{C}_{ij}$ represents the covariance matrix between $d_{i}$ and $d_{j}$.  The distribution of $d_2$ conditioned on $d_1$ is then also a multivariate Gaussian given by
    \begin{equation}
    P(d_2 | d_1) = \mathcal{N} \qty( \mu_2 + \mathbf{C}_{21}^{-1} \mathbf{C}_{11}^{-1} (d_1 - \mu_1), \, \mathbf{C}_{22} - \mathbf{C}_{21}\mathbf{C}_{11}^{-1} \mathbf{C}_{12} ).
    \end{equation}
    We will use this expression when computing the PPD for disjoint subsets of the full \3x2pt data vector.
\end{itemize}

\subsubsection{Calibration of PPD tests applied to DES Y1}
\label{sec:pval_calib}

Given the potential for $p$-values to be biased low for consistency tests where $d_{\rm obs}$ and $d_{\rm rep}$ represent different observables (or other splits of the full data vector), we wish to be able to tell whether an observed $p$-value is biased low due to the effect described in \cref{sec:low_pvalues} or if it is rather indicative of an actual tension in the data. To do so, we will attempt to calibrate $p$-values by sampling the distribution of such $p$-values for simulated data vectors which we know to be generated consistently. Given a fiducial cosmology (taken here as the best-fit of the full \3x2pt analysis), we generate noisy data vectors by sampling the Gaussian likelihood and measure the $p$-values for the same PPD tests as those applied to real data. The comparison of the distribution of $p$-values for simulated data vectors to the $p$-value for observed data will provide a {\it calibrated} $\tilde{p}$-value, which is the fraction of simulated data vectors yielding a lower $p$-value than the observed data.

It would be prohibitively expensive in computing time to run a Markov chain for each simulated (noisy) data vector. Therefore, for each PPD test, we instead run a single chain on the fiducial (noiseless) data vector and use importance sampling to reweight samples in the chain for each simulated data vector, before recomputing PPD test statistics. The successive steps are the following.
\begin{enumerate}
    \item We run a standard Markov chain for the fiducial data vector,~$d_{\rm fid}$, to generate a sample of parameters $\Theta_i \sim P(\Theta|d_{\rm fid})$ with weights $w_i$ (see \cref{sec:sampling}).
    \item For each simulated data vector $d_{{\rm sim},j}$, we repeat the following steps\footnote{Note that, since the likelihood is Gaussian, these steps amount to recomputing various differences of log-likelihoods, which can be easily parallelized and performed within a few minutes for a sample of $10^5$ simulated data vectors.}:
    \begin{enumerate}
        \item We compute the importance weights, given by the ratios of the posteriors for simulated and fiducial data vectors, ${\alpha_{ij} = P(d_{{\rm sim},j}|\Theta_i) / P(d_{\rm fid}|\Theta_i)}$, and multiply them by the original weights $w_i$ to get updated weights $w_{ij}=w_i\alpha_{ij}$;
        \item We compute the PPD test statistics $T(d_{{\rm sim},j},\Theta_i)$;
        \item We draw samples from the PPD by generating a realization $d_{{\rm rep},i}$ at each parameter sample $\Theta_i$ (conditioned on $d_{{\rm sim},j}$ if calibrating a consistency test), and compute test statistics $T(d_{{\rm rep},i}|\Theta_i)$;
        \item We compute the $p$-value, $p_j$, given statistics $T(d_{{\rm sim},j},\Theta_i)$ and $T(d_{{\rm rep},i}|\Theta_i)$ with weights $w_{ij}$ using \cref{eq:pvalue}.
    \end{enumerate}
    \item We use the distribution of $p_j$ obtained from $N_{\rm sim}=10^5$ independent simulated data vectors to calculate the calibrated $\tilde{p}$-value for a given test with $p$-value $p$, such that
    \begin{equation}
        \tilde{p} = \frac{1}{N_{\rm sim}} \sum_{j} \mathbb{1}(p-p_j),
    \end{equation}
    where $\mathbb{1}$ is the Heaviside function.
\end{enumerate}
In practice, we find that, given our sampling strategy (see \cref{sec:sampling}), the importance sampling procedure results in relatively low errors on the estimated (raw) $p$-values, based on effective number of samples (of order few hundreds), thus validating the method.
We will therefore report calibrated $\tilde{p}$-values for each test and replace our consistency criterion by $\tilde{p}>0.01$.

Finally, we mention that we will consider many different tests of the data below.  By performing multiple tests, we are necessarily more likely to obtain evidence for tension by random chance.  One method to correct for this effect is the Bonferroni correction \citep{Dunn:1959}, which scales down the $p$-value threshold by a factor equal to the number of hypothesis tests.  However, this correction can be overly severe, particularly when the data are correlated (as is the case here).  Since the number of goodness-of-fit tests that we apply is fairly small (essentially only four), we will generally ignore corrections to our $p$-value threshold for multiple hypothesis tests.  In some sense, this is a conservative approach since it makes us more likely to be worried about possible tensions.  While we report many $p$-values for individual redshift bins (and more) below, we will not consider the data to be in internal tension if a few of these $p$-values fall below our $p$-value threshold.  Instead, we will take the approach of considering subsets of the data with low $p$-values to warrant more exploration.  In other words, we are not worried about the possibility of type-I errors, where a true null hypothesis is rejected, since we mainly use these $p$-values as a means to investigate agreement between the model and data.

\subsection{Sampling methodology}
\label{sec:sampling}

As described above, we generate PPD realizations from a posterior by drawing from the (possibly conditional) likelihood at each point in a posterior chain.  The posterior chains are generated using \texttt{PolyChord} \citep{Handley:2015fda, Handley:2015}. \texttt{PolyChord} uses Nested Sampling to calculate both the Bayesian evidence and the posterior distribution. It overcomes some of the issues of ellipsoidal nested sampling by using slice sampling \citep{Neal:2000, Aitken:2013}. This makes it the optimal sampler for the DES likelihood, as discussed in \cite{Samplers}.

As described in \citet{3x2pt}, we sample over the standard \lcdm parameter space, combined with the parameters describing systematic errors in the DES data.  The DES-specific parameters describe multiplicative shear biases in the shear measurements, redshift biases in the assumed source and lens galaxy redshift distributions, and linear galaxy bias parameters.  We additionally allow the neutrino mass to vary.  Details of the assumed priors can be found in \citet{3x2pt}.

Once the parameter chain for a given posterior is generated, we re-process the chain using \texttt{CosmoSIS} \citep{Zuntz:2015} to generate the realizations from the likelihood at each step in the chain.  Since \texttt{PolyChord} generates weighted samples of the posterior, these weights are also applied to all PPD realizations (and multiplied by importance weights for calibration purposes).

\section{Application of Posterior Predictive Checks to DES Y1 Data}\label{sec:application}

We now apply the PPD formalism developed above to various splits of the DES Y1 measurements.  We consider several splits, motivated by considerations of potential systematic errors and possible beyond \lcdm physics.  

For visual clarity, when plotting the data we subtract the best-fit \3x2pt model from both the data and the PPD realizations, and normalize relative to the diagonals of the covariance matrix.  In other words, we plot
\begin{equation}
\delta X_i = \frac{d_i - \mu^{\rm MAP}_i}{\sqrt{\mathbf{C}_{ii}}},
\end{equation}
where $d_i$ is either the true data or the PPD realizations of the data,  $\mu^{\rm MAP}_i$  is the best fit to the {\it full} \3x2pt data vector (with fiducial scale cuts), and $\mathbf{C}$ is the covariance matrix of the \3x2pt data vector.  The choice to plot $\delta X_i$ rather than $d_i$ has no impact on the comparison between data and realizations and makes the plots easier to visualize.

\Cref{tab:pvalues} summarizes $\tilde{p}$-values for the full data vector as well as each individual probe for each test considered in the following \crefrange{sec:3x2_goodness_of_fit}{sec:xim_xip}.

\setlength{\tabcolsep}{2pt}
\begin{table*}
    \centering
    \begin{tabular}{p{4cm} c c c c c c c}
        \toprule
         & & & \multicolumn{5}{c}{PPD test calibrated $\tilde{p}$-values} \\
        \cmidrule(l{2pt}r{2pt}){4-8} 
        Test & $d_{\rm rep}$ & $d_{\rm obs}$ & $d_{\rm rep} | d_{\rm obs}$ & $\xip | d_{\rm obs}$ & $\xim | d_{\rm obs}$ & $\gammat | d_{\rm obs}$ & $w | d_{\rm obs}$ \\
        \midrule
        \textit{Goodness-of-fit tests}  & & &  &  &  &  &  \\
        Full 3x2pt & $\qty{\xip,\xim,\gammat,w}$ & $\qty{\xip,\xim,\gammat,w}$ & 0.065 (0.046) & 0.537 & 0.182 & 0.238 & 0.071 \\
        Cosmic shear & $\qty{\xip,\xim}$ & $\qty{\xip,\xim}$ & 0.386 (0.396) & 0.533 & 0.192 & -- & -- \\
        Galaxy-galaxy lensing & $\qty{\gammat}$ & $\qty{\gammat}$ & 0.262 (0.245) & -- & -- & 0.262 & -- \\
        Clustering & $\qty{w}$ & $\qty{w}$ & 0.057 (0.050) & -- & -- & -- & 0.057 \\
        \midrule
        \textit{Consistency tests: data type splits} & & & & & & & \\
        Cosmic shear \vs galaxy-galaxy lensing and clustering & $\qty{\xip,\xim}$ & $\qty{\gammat,w}$ &  0.396 (0.299) & 0.600 & 0.162 & -- & -- \\
        Galaxy-galaxy lensing \vs cosmic shear and clustering & $\qty{\gammat}$ & $\qty{\xip,\xim,w}$ & 0.336 (0.142) & -- & -- & 0.336 & -- \\
        Clustering \vs cosmic shear and galaxy-galaxy lensing & $\qty{w}$ & $\qty{\xip,\xim,\gammat}$ & 0.050 ($3.6\times10^{-5}$) & -- & -- & -- & 0.050 \\
        \midrule
        \textit{Consistency tests: other} & & & & & & & \\
        Cosmic shear: bin 1 \vs no bin 1 & $\qty{\xip,\xim}$ & $\qty{\xip,\xim}$ & 0.639 (0.532) & 0.648 & 0.543 & -- & -- \\
        Cosmic shear: bin 2 \vs no bin 2 & $\qty{\xip,\xim}$ & $\qty{\xip,\xim}$ & 0.392 (0.344) & 0.379 & 0.366 & -- & -- \\
        Cosmic shear: bin 3 \vs no bin 3 & $\qty{\xip,\xim}$ & $\qty{\xip,\xim}$ & 0.547 (0.372) & 0.771 & 0.287 & -- & -- \\
        Cosmic shear: bin 4 \vs no bin 4 & $\qty{\xip,\xim}$ & $\qty{\xip,\xim}$ & 0.376 (0.293) & 0.593 & 0.095 & -- & -- \\
        \3x2pt: large \vs small scales & $\qty{\xip,\xim,\gammat,w}$ & $\qty{\xip,\xim,\gammat,w}$ & 0.034 (0.016) & 0.091 & 0.741 & 0.167 & 0.030\\
        Cosmic shear: $\xim$ \vs $\xip$ & $\qty{\xim}$ & $\qty{\xip}$ & 0.186 (0.151) & -- & 0.186 & -- & --  \\
        \bottomrule
    \end{tabular}
    \caption{Summary of calibrated $\tilde{p}$-values obtained for all tests.  The ``Test'' column specifies the test.  The second and third columns show the observables considered for sampling ($d_{\rm rep}$) and conditioning ($d_{\rm obs}$). The fourth column shows the $\tilde{p}$-value for all observables considered in the test with the uncalibrated $p$-value in parenthesis, while the rest of the columns show the $\tilde{p}$-value when considering individual observables. 
    }
    \label{tab:pvalues}
\end{table*}

\subsection{Goodness of fit of the full \3x2pt data vector}
\label{sec:3x2_goodness_of_fit}

\begin{figure*}
    \centering
    3x2pt goodness of fit \\
    \begin{tikzpicture}
    \node(a){\includegraphics[trim=0 0 0 0.5cm, scale=0.4]{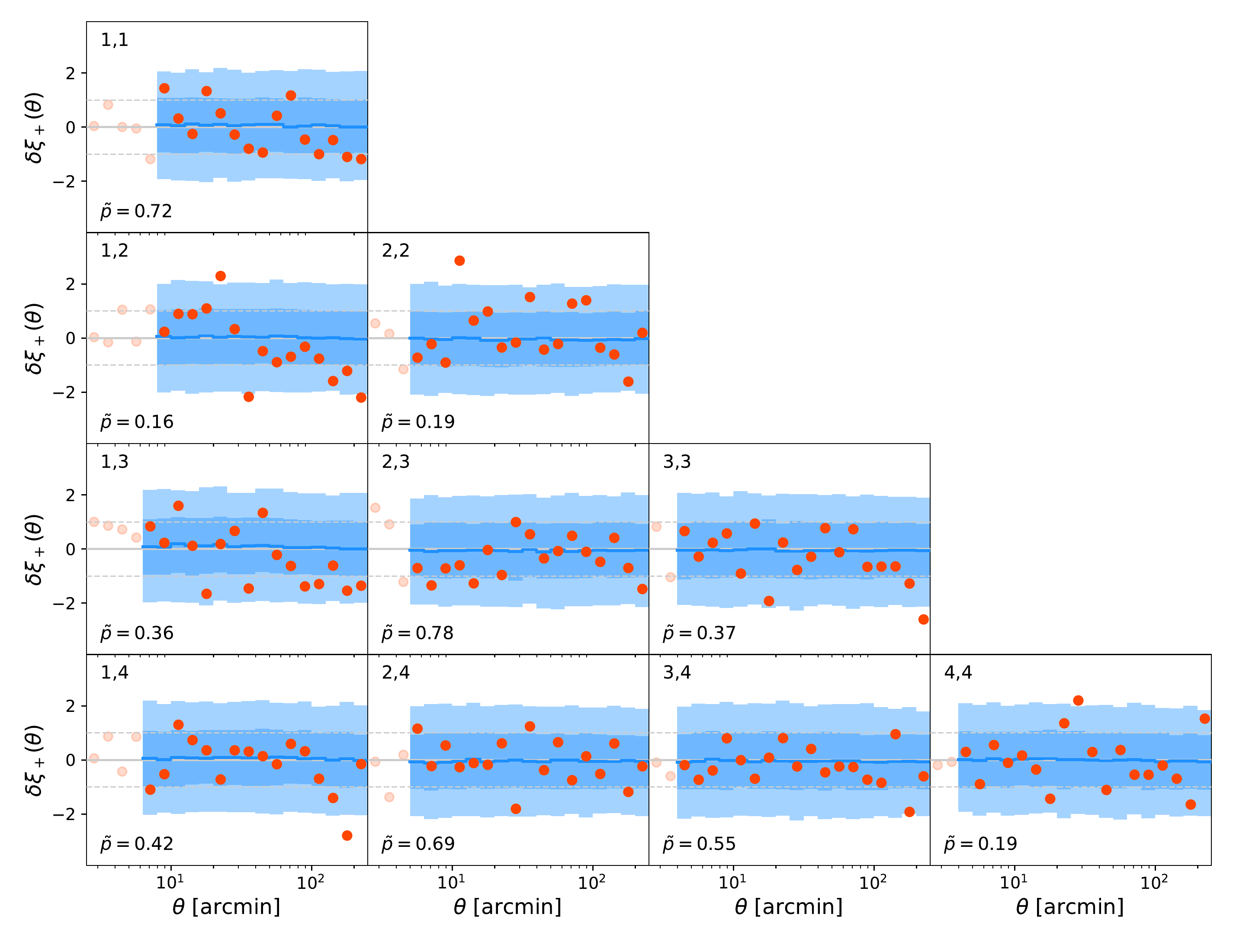}};
    \node at (a.north east)
    [
    anchor=center,
    xshift=-25mm,
    yshift=-35mm
    ]
    {
        \includegraphics[trim=0 0 0 0.5cm, scale=0.4]{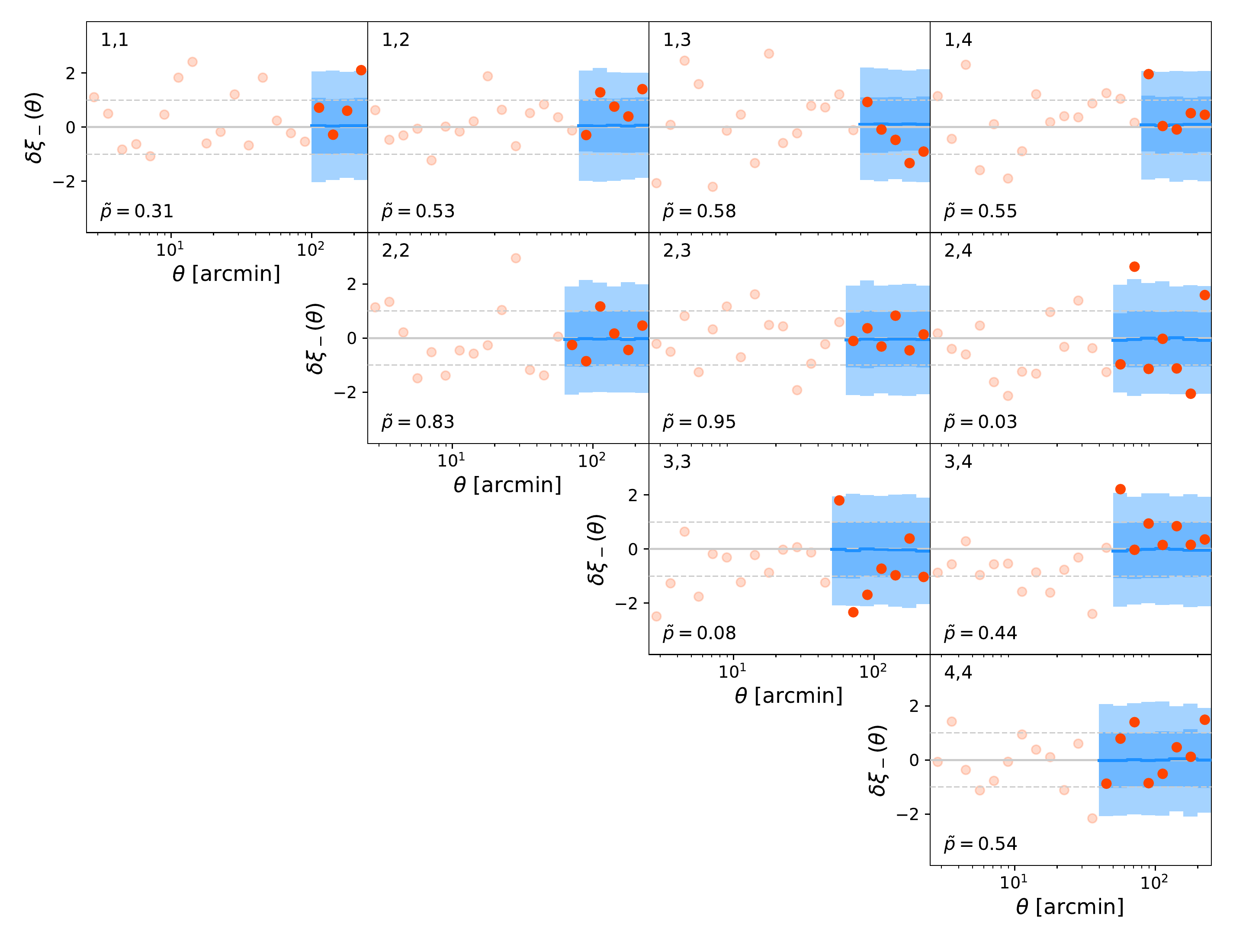}
    };
    \end{tikzpicture}
    \includegraphics[trim=0 0 0 0.5cm, scale=0.4]{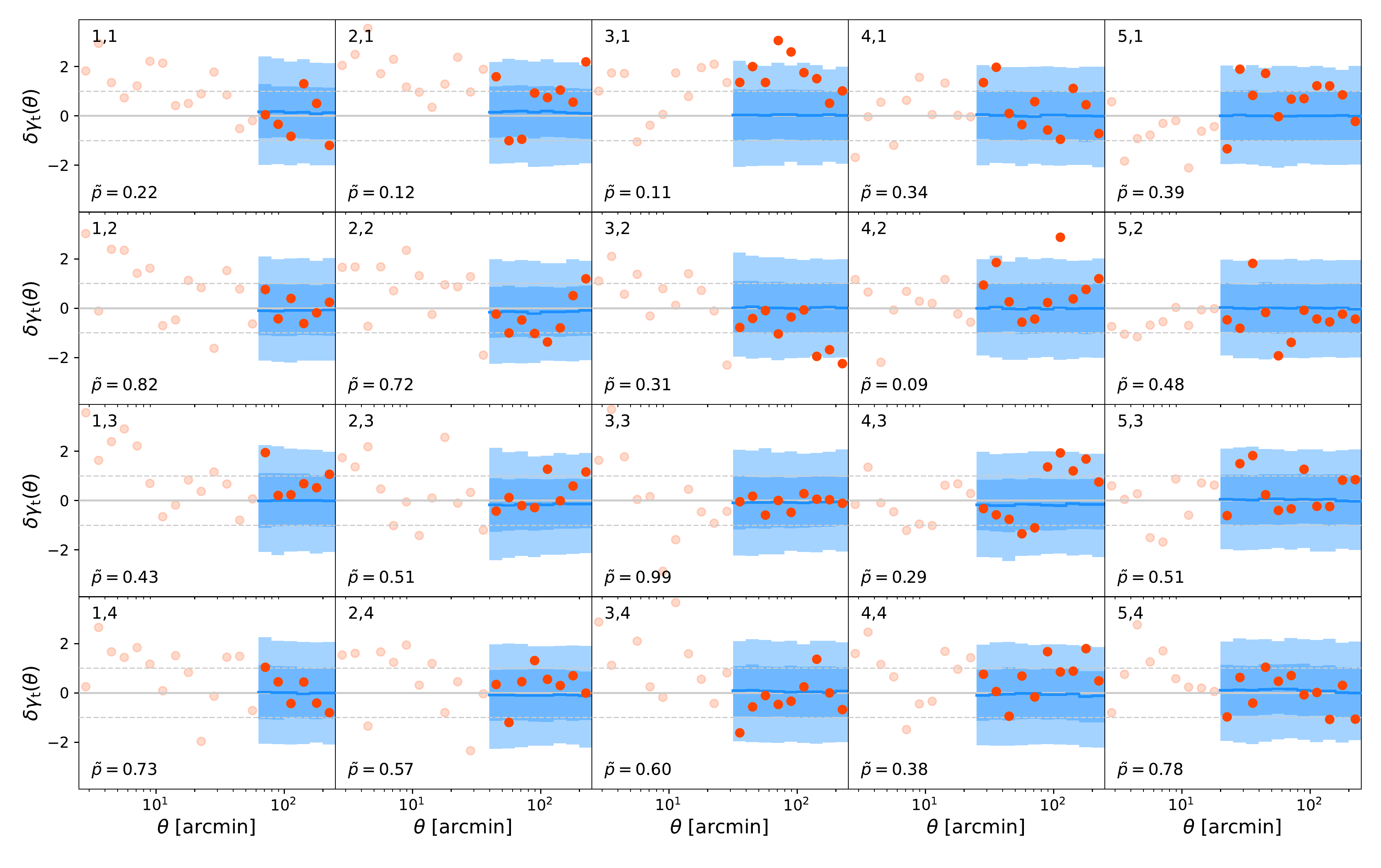}
    \includegraphics[trim=0 0 0 0.5cm, scale=0.4]{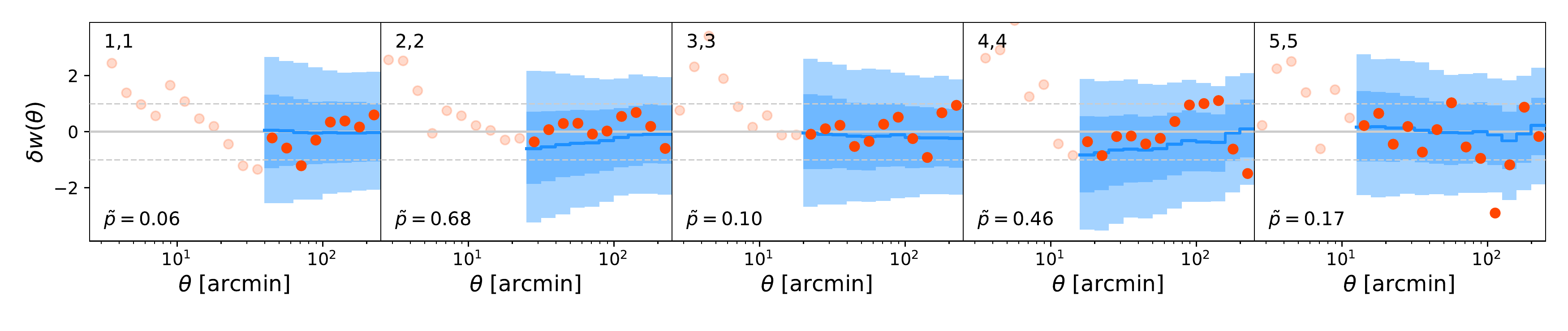}
    \caption{PPD goodness-of-fit test for the \3x2pt data vector.  The 68\% and 95\% confidence bands on the PPD realizations of \3x2pt, conditioned on the posterior from the analysis of the \3x2pt data is shown as the blue bands.  Red points represent the actual data.
    The bands and the data points are plotted relative to the best-fit \3x2pt theory curve, and are normalized by the diagonal of the \3x2pt covariance, such that data error bars have unit size.
    The different insets split the datavector into the different observables ($\xipm$, $\gammat$, and $\wtheta$), while the different panels split the correlation functions by redshift bin combination.
    The calibrated $\tilde{p}$-value for each redshift bin combination of each observable is indicated in the bottom left corner of the corresponding panel and calibrated $\tilde{p}$-values per observable and for the entire set are reported in \cref{tab:pvalues}. \label{fig:ppd_3x2}}
\end{figure*}

We first consider the PPD for the full \3x2pt data vector, shown in \cref{fig:ppd_3x2}.  This first test is useful to determine whether the model favored by the data is actually a good fit to the data. However, unlike the classical $\chi^2$ test which only uses the best-fit model, the PPD goodness-of-fit test marginalizes over model parameter uncertainties.

The different insets in \cref{fig:ppd_3x2} split the datavector into the different observables ($\xipm$, $\gammat$, and $\wtheta$), while the different panels split the correlation functions by redshift bin combination.  The distribution of PPD  realizations is shown as the blue bands, while the actual data is shown as the red points. The faded out points represent measurements that were not included because of angular scale cuts.

As discussed in \cref{sec:ppd}, the computation of a $p$-value using the test statistic of \cref{eq:test_stat} is motivated for goodness-of-fit tests like that considered in \cref{fig:ppd_3x2}.
We also perform the calibration test, repeating the same PPD goodness-of-fit test for $10^5$ simulated, noisy data vectors sampled at the \3x2pt best-fit cosmology, and using importance sampling to rapidly compute corresponding $p$-values. We show their distribution in \cref{fig:meta_pval}. We then compute a calibrated $\tilde{p}$-value, given by the ratio of simulated $p$-values below the one measured from data. We find similar values for this specific test, which indicates that, as expected, our choice of statistics has little impact on goodness-of-fit tests.

We compute a calibrated $\tilde{p}$-value of 0.065 ($p = 0.046$ uncalibrated) for the full \3x2pt data vector, indicating an acceptable fit given our criterion of $\tilde{p}>0.01$. We report calibrated $\tilde{p}$-values for individual probes in \cref{tab:pvalues}. Additionally, we find calibrated $\tilde{p}$-values of 0.373 when considering both components of cosmic shear, and 0.111 when considering $\gammat$ and $\wtheta$.
Each panel of \cref{fig:ppd_3x2} also indicates the calibrated $\tilde{p}$-value computed for that particular subset of the data (still using the PPD realization from the full \3x2pt data vector). Given the potential pitfalls of multiple hypothesis testing, we use these individual $\tilde{p}$-values mostly to rank the different bin combinations by largest discrepancy with the model. 
The lowest $\tilde{p}$-value is obtained for the $(2,4)$ redshift bin combination of $\xim$. We discuss this particular subset of the data more in \cref{sec:bad_xim}.

\begin{figure*}
\begin{center}
Clustering \vs cosmic shear and galaxy-galaxy lensing
\includegraphics[scale=0.4]{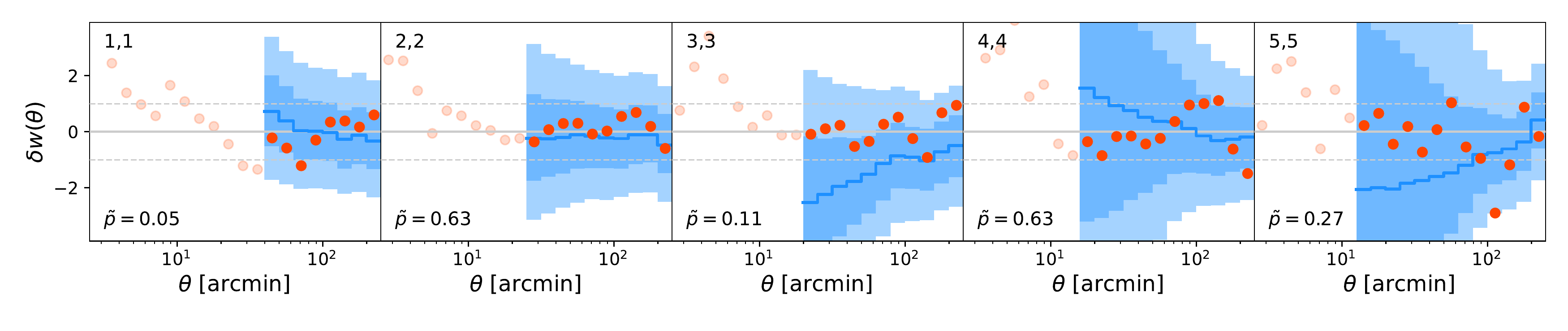}
\caption{PPD for clustering, conditioned on the posterior from cosmic shear and galaxy-galaxy lensing.  See \cref{fig:ppd_3x2} for explanation of bands. \label{fig:ppd_wtheta_vs_gggammat}}
\end{center}
\end{figure*}

\subsection{Goodness of fit of individual two-point functions}

We now consider goodness-of-fit tests of the different two-point function components of the full \3x2pt data vector separately, i.e. we test for the goodness-of-fit of cosmic shear $\xipm$, galaxy-galaxy lensing $\gammat$ and clustering $w$, one at a time.  This differs from the test considered in \cref{sec:3x2_goodness_of_fit} in that we condition on the posteriors for each subset of the \3x2pt data vector, rather than on the posterior from the analysis of the full \3x2pt vector. This allows us, first, to test each probe separately, and then to exclude the case where the \3x2pt test $\tilde{p}$-value is dominated by one or two probes, which could mask a poor fit to the third.  Given the similarities between these tests and that of \cref{sec:3x2_goodness_of_fit}, we relegate the associated plots to \cref{app:goodness_of_fit}.

\Cref{fig:gg_goodness_of_fit} shows the graphical PPD test for cosmic shear.  There are no obvious discrepancies between the PPD realizations of cosmic shear and the actual data.  
We compute a calibrated $\tilde{p}$-value for cosmic shear of $\tilde{p}=0.386$, indicating no evidence for tension between the measurements and PPD realizations.  As for the \3x2pt goodness-of-fit test, though, the $(2,4)$  bin combination of $\xim$ shows a low $p$-value with $\tilde{p}=0.013$.
\Cref{fig:gammat_goodness_of_fit} shows the PPD realizations for galaxy-galaxy lensing.  The PPD realizations in this case look generally consistent with the data.  Considered as a whole, though, the $\gammat$ measurements exhibit a good fit, with $\tilde{p}=0.262$.
\Cref{fig:wtheta_goodness_of_fit} shows the PPD realizations for clustering.  Again, the realizations show good agreement with the data.  The $p$-value for $\wtheta$ is $\tilde{p}=0.057$.

\subsection{Testing for internal consistency between the two-point functions}
\label{sec:1x2pt_vs_2x2pt}

We next consider PPD realizations of the form $P(d_1^{\rm rep} \, |\, d_2^{\rm obs})$, where $d_1$ represents one of $\qty{ \xipm, \gammat, w }$ and $d_2$ represents the remaining elements of the \3x2pt data vector from the two other probes.  Such tests are interesting for several reasons.  First, tests of this form can be used to split parts of the \3x2pt data vector that depend on different known systematics, such as shear calibration bias.
For example, additive shear systematics can impact measurement of $\xipm$, but are expected to not impact $\gammat$ significantly, and have no impact on clustering. Therefore, if we set $d_1 = \qty{w}$ and $d_2 = \qty{\xipm,\gammat}$, then $d_2$ is impacted by potential shear biases while $d_1$ is not.
Similarly, if $d_1 = \qty{\xipm}$ while $d_2 = \qty{ \gammat, w }$, then $d_2$ will be impacted by potential biases in the lens galaxy redshifts, while $d_1$ will not.  Secondly, departures from \lcdm might be expected to appear in some observables, but not in others.  For instance, a split for which $d_2$ depends on lensing while $d_1$ does not would show tension in models where lensing is impacted by beyond-\lcdm physics while clustering is not.  For example, this is the case for the $(\Sigma,\mu)$ parametrizations of modified gravity that perturb the Poisson equation differently for light and matter, as explored with DES Y1 data \citep{2019PhRvD..99l3505A} and other galaxy surveys and CMB data \citep{2013MNRAS.429.2249S,2018MNRAS.475.2122M,2019PhRvD..99h3512F,2020A&A...641A...6P}.

We note that the consistency  tests considered in this section are similar to the test of consistency between $\xipm$ and $\{\gammat, w \}$ considered in \citet{3x2pt}, where these two subsets of the \3x2pt data vector were found to be consistent.  However,  the test presented in \citet{3x2pt} used an evidence ratio to identify consistency.
Furthermore,  the evidence ratio test naturally lives in model space rather than data space.  Consequently,  it is difficult to use the evidence ratio test to evaluate particularly discrepant elements of the data vector.  The PPD test on the other hand, does this naturally.

\Cref{fig:ppd_wtheta_vs_gggammat} shows the PPD realizations for clustering, conditioned on the observed cosmic shear and galaxy-galaxy lensing measurements.  Observational systematics, such as galactic dust, are expected to impact clustering more than cosmic shear and galaxy-galaxy lensing.  Galactic dust would tend to obscure galaxies, modulating the number density (and thus $\wtheta$), but likely not having a large impact on inferred shears.  Furthermore,  as pointed out by \citet{Leauthaud:2017}, gravitational lensing measurements around BOSS-selected galaxies show a lower amplitude signal than expected based on the clustering properties of the galaxies.  If a similar discrepancy were born out in DES data, this test would be expected to reveal it. 

The results in \cref{fig:ppd_wtheta_vs_gggammat} show that $\wtheta$ appears generally within the bounds of the PPD realizations.
Given the large covariance between the $\wtheta$ points, this appearance can be deceiving: we find that the uncalibrated $p$-values for several of the $\wtheta$ redshift bins are quite low, in the range of $0.01$ to $0.02$. However, we report a calibrated $\tilde{p}$-value of 0.050, revealing agreement between clustering measurements and expectations from shear and galaxy-galaxy lensing measurements. We therefore note that this test is an example of the case presented in \cref{sec:low_pvalues}, where the part of the data vector that is being tested has a different parameter dependence than the part that is used for conditioning. Namely, $\wtheta$ is more sensitive to linear galaxy bias than $\{\gammat,\xipm\}$ and exhibits different parameter degeneracies.  The theory data vectors and consequently PPD realizations for $\wtheta$ conditioned on $\{\gammat,\xipm\}$ therefore have amplitudes spread over a wide range, which is visible in \cref{fig:ppd_wtheta_vs_gggammat}. Consequently, the uncalibrated $p$-value in this case will be driven low as discussed in \cref{sec:low_pvalues} and shown in \cref{fig:meta_pval}.

We relegate the two other PPD comparisons of this type to \cref{app:conditioned}.  \Cref{fig:ppd_gammat_vs_ggwtheta} shows the PPD realizations for galaxy-galaxy lensing, conditioned on the observed clustering and cosmic shear measurements.   In this case, we find that all of the data points appear to be quite consistent with the PPD realizations, and the $\tilde{p}$-values all appear to be reasonable, with an overall $\tilde{p}$-value of \num{0.336}.
\Cref{fig:ppd_gg_vs_wthetagammat} shows the PPD realizations for cosmic shear, conditioned on the observed clustering and galaxy-galaxy lensing measurements.
Nevertheless we note that there appears to be a weak trend for the $\xip$ data points at large scales to be low relative to the PPD realizations. 
Furthermore, as seen previously, the $(2,4)$ bin of $\xim$ yields a low $p$-value.
We find in this case that most of the data appear reasonable given the PPD realizations, with an overall $\tilde{p}$-value of \num{0.396}.

\subsection{Testing for cosmic shear redshift-dependent inconsistency}

\begin{figure*}
    \begin{center}
    Cosmic shear: bin 1 \vs no bin 1\\
    \includegraphics[scale=0.4, trim={0 0.8cm 0 0},clip]{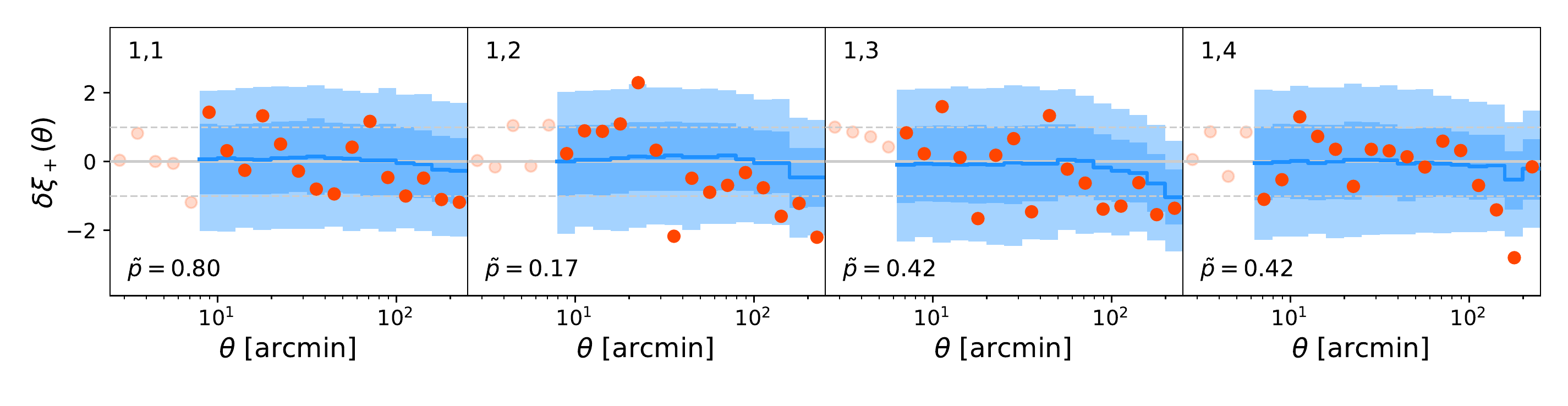}
    \includegraphics[scale=0.4, trim={0 0.8cm 0 0},clip]{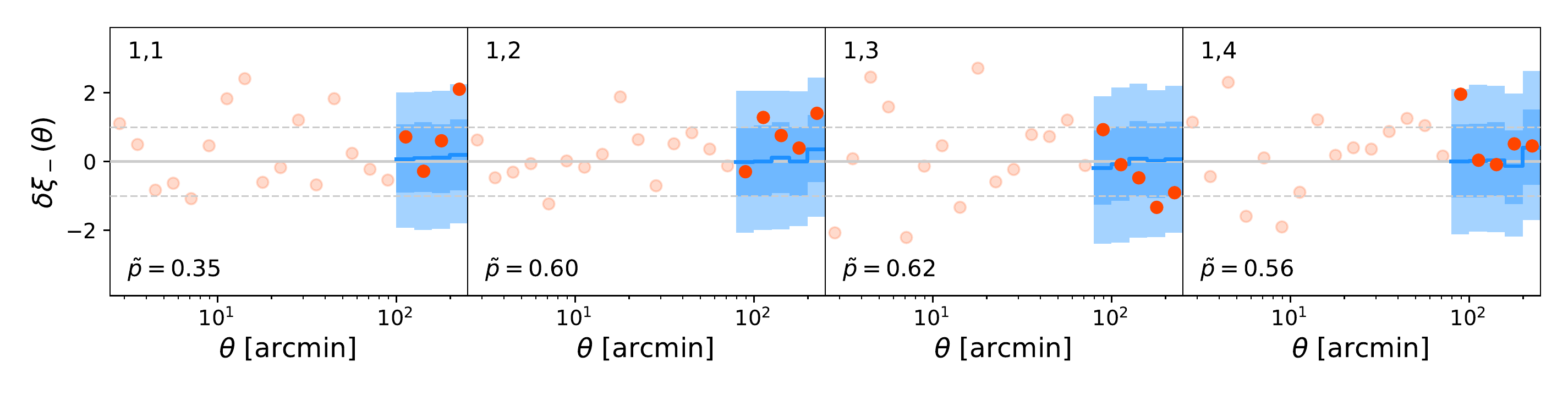}
    
    Cosmic shear: bin 2 \vs no bin 2\\
    \includegraphics[scale=0.4, trim={0 0.8cm 0 0},clip]{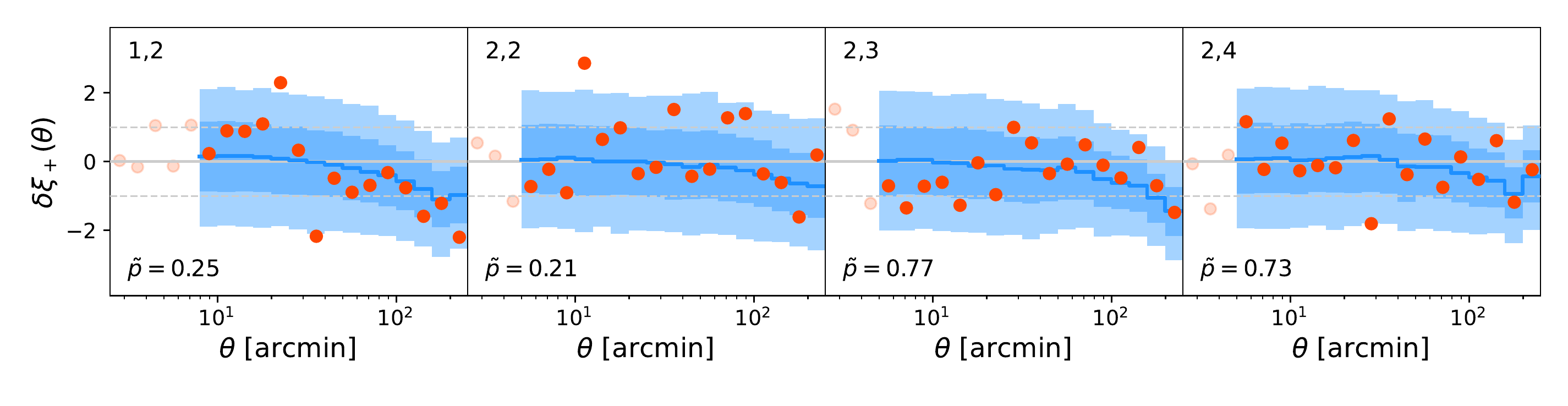}
    \includegraphics[scale=0.4, trim={0 0.8cm 0 0},clip]{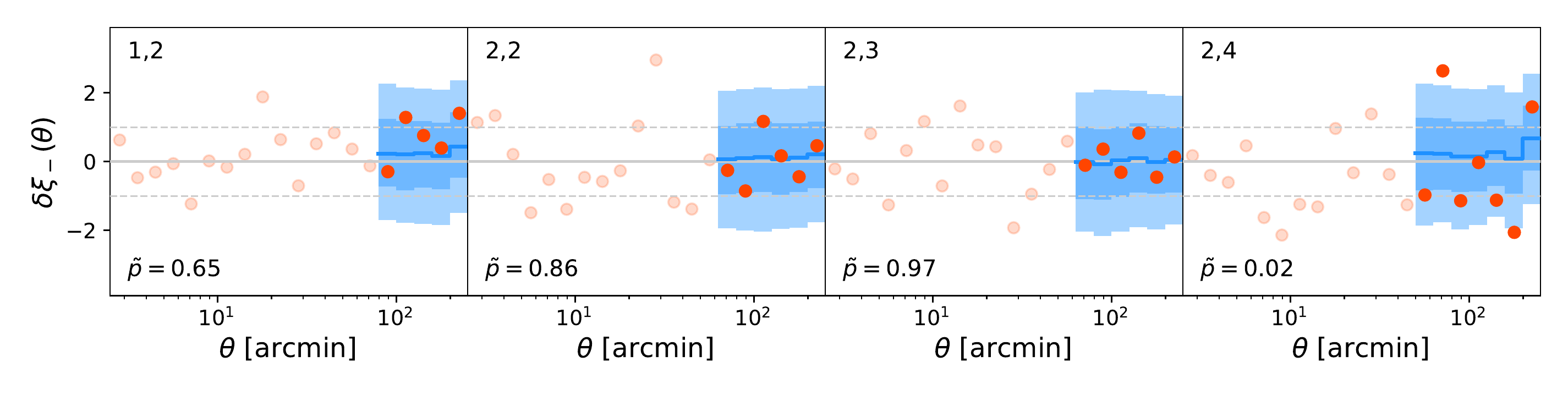}
    
    Cosmic shear: bin 3 \vs no bin 3\\
    \includegraphics[scale=0.4, trim={0 0.8cm 0 0},clip]{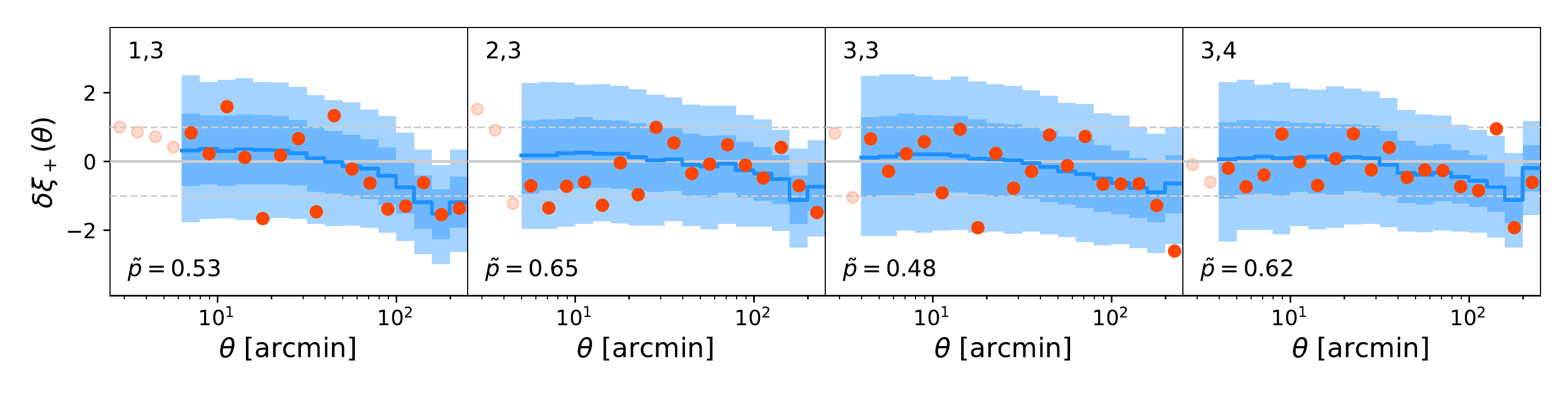}
    \includegraphics[scale=0.4, trim={0 0.8cm 0 0},clip]{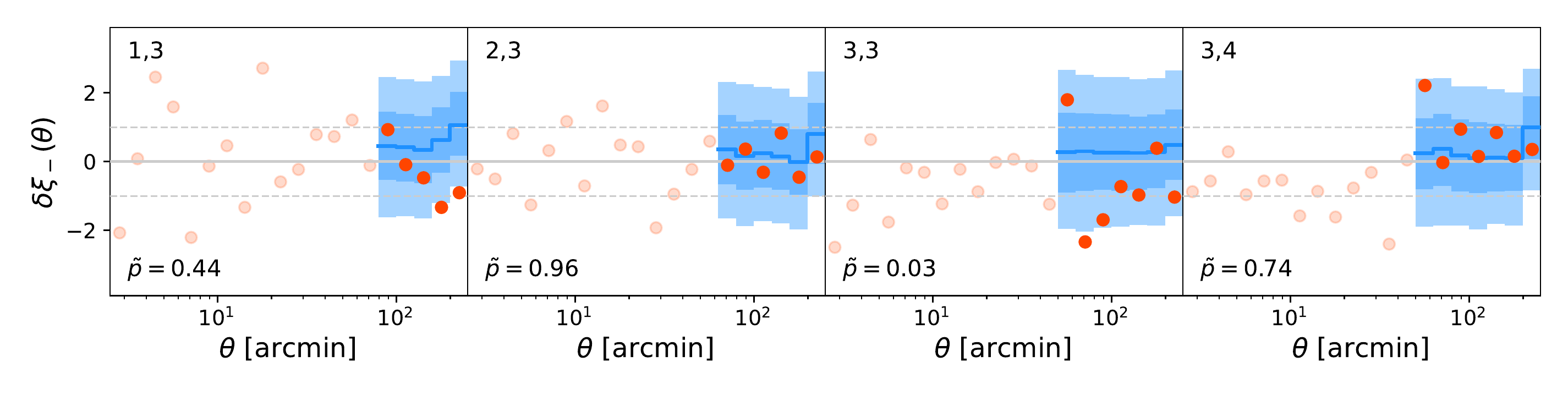}
    
    Cosmic shear: bin 4 \vs no bin 4\\
    \includegraphics[scale=0.4, trim={0 0.8cm 0 0},clip]{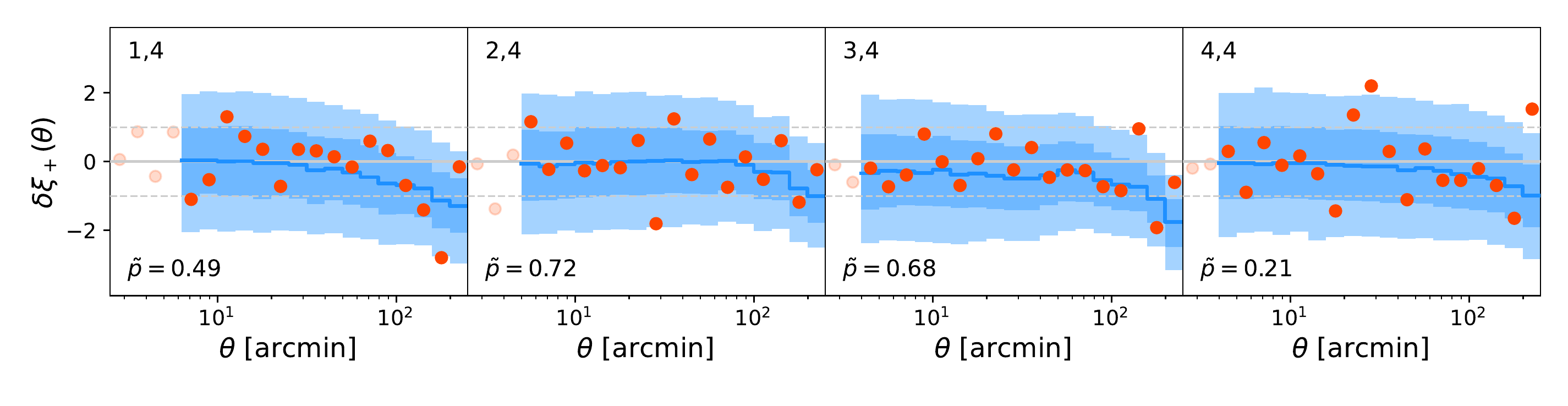}
    \includegraphics[scale=0.4, trim={0 0.8cm 0 0},clip]{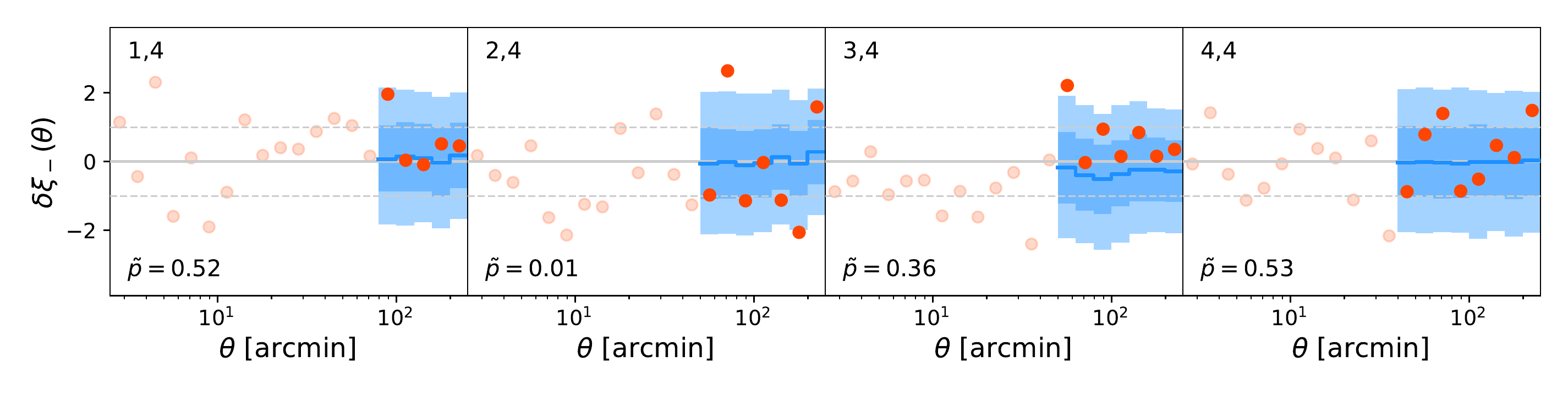}
    \end{center}
    \caption{PPD for cosmic shear where one redshift bin is tested at a time, conditional to the posterior obtained from cosmic shear by removing auto- and cross-correlations with that bin.   See \cref{fig:ppd_3x2} for explanation of bands. \label{fig:gg_zbins}}
\end{figure*}

Splits of the data by redshift bin are motivated both as a probe of departures from \lcdm and by concerns of systematic errors.  Models with an evolving dark energy equation of state, for instance, would predict redshift-dependent departures from \lcdm that could be revealed by such tests.  Similarly, shear measurements at high redshift could be more impacted by issues such as source blending and PSF modeling uncertainty, while low-redshift measurements are more impacted by modeling errors of the non-linear matter power spectrum and intrinsic alignments, which might then lead to tension between low and high-redshift measurements.

One complication of splitting the data on redshift bin is that certain model parameters---such as galaxy bias or the multiplicative shear bias parameters---impact only one of the redshift bins.  These parameters will then be unconstrained when conditioning on the other redshift bins.  For such  parameters, the PPD realizations will then involve drawing from the parameter priors.  This is not problematic for parameters like multiplicative shear bias and photometric redshift bias, which are prior dominated anyways, but is problematic for linear galaxy bias.  If a bias parameter is unconstrained, the PPD realizations will necessarily span a much broader range than the data, making the graphical PPD tests difficult.  We avoid this issue  by focusing on redshift splits of the cosmic shear data vector, which  is unaffected by galaxy bias.

\Cref{fig:gg_zbins} shows the PPD realizations for single bins of cosmic shear, conditioned on the realizations from the other redshift bins. For instance, the upper plot shows the PPD consistency test for bins (1,1), (1,2), (1,3) and (1,4), conditioned on measurements of all other redshift bin combinations. The PPD realizations and $\tilde{p}$-values generally appear reasonable, although again the $(2,4)$ bins of $\xim$ consistently exhibits a $\tilde{p}$-value close to 0.01. However, all overall calibrated $\tilde{p}$-values are well above 0.01, indicating no sign of tensions between redshift bins in DES Y1 cosmic shear data.

\subsection{Testing for large and small-scale systematics}
\label{sec:3x2_largescales_vs_3x2_smallscales}

\begin{figure*}
    \begin{center}
    3x2pt large scales ($\theta>\SI{100}{\arcminute}$) vs small scales ($\theta<\SI{100}{\arcminute}$) \\
    \begin{tikzpicture}
    \node(a){\includegraphics[scale=0.4]{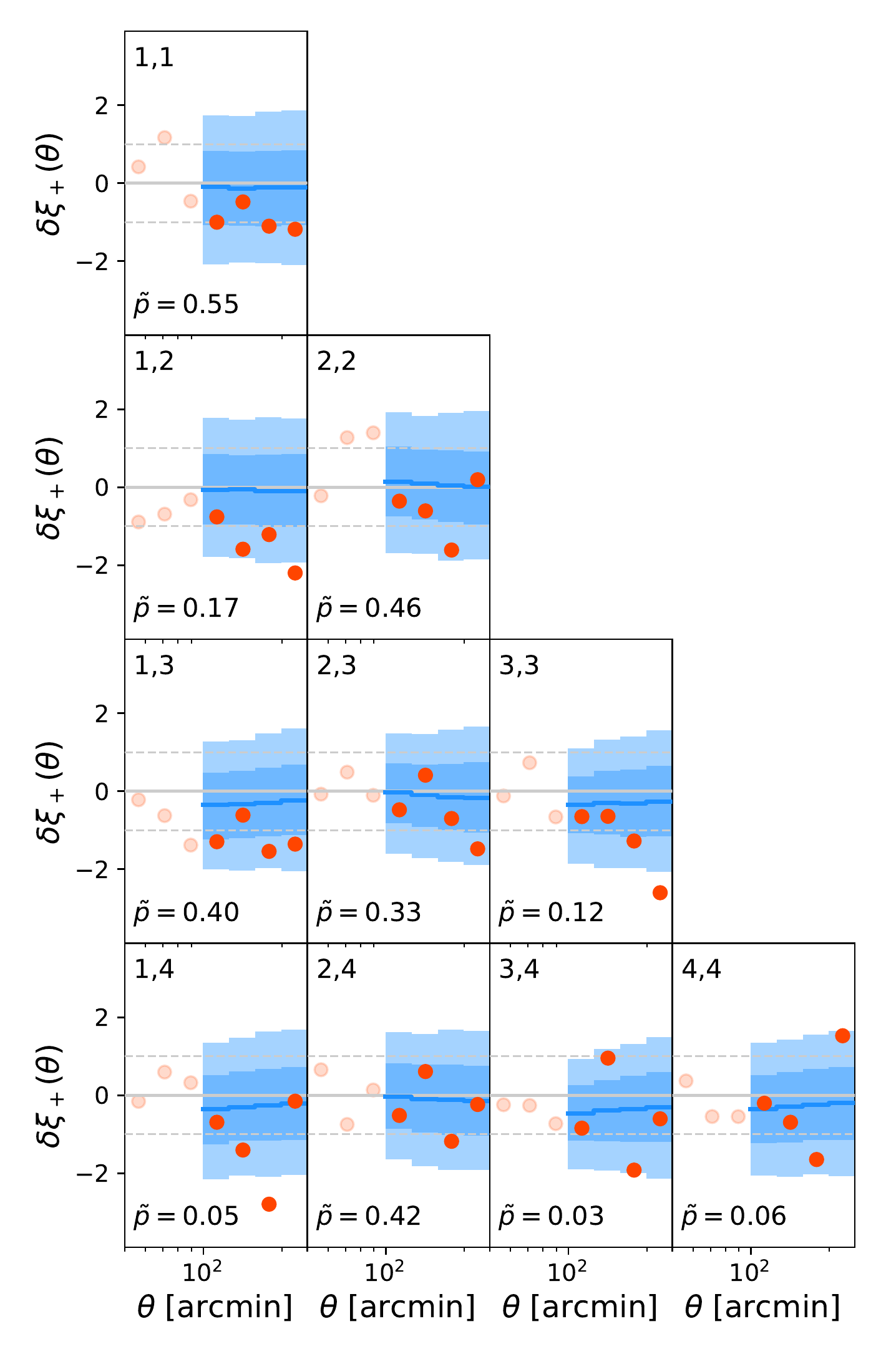}};
    \node at (a.east)
    [
    anchor=center,
    xshift=0mm,
    yshift=0mm
    ]
    {
        \includegraphics[scale=0.4]{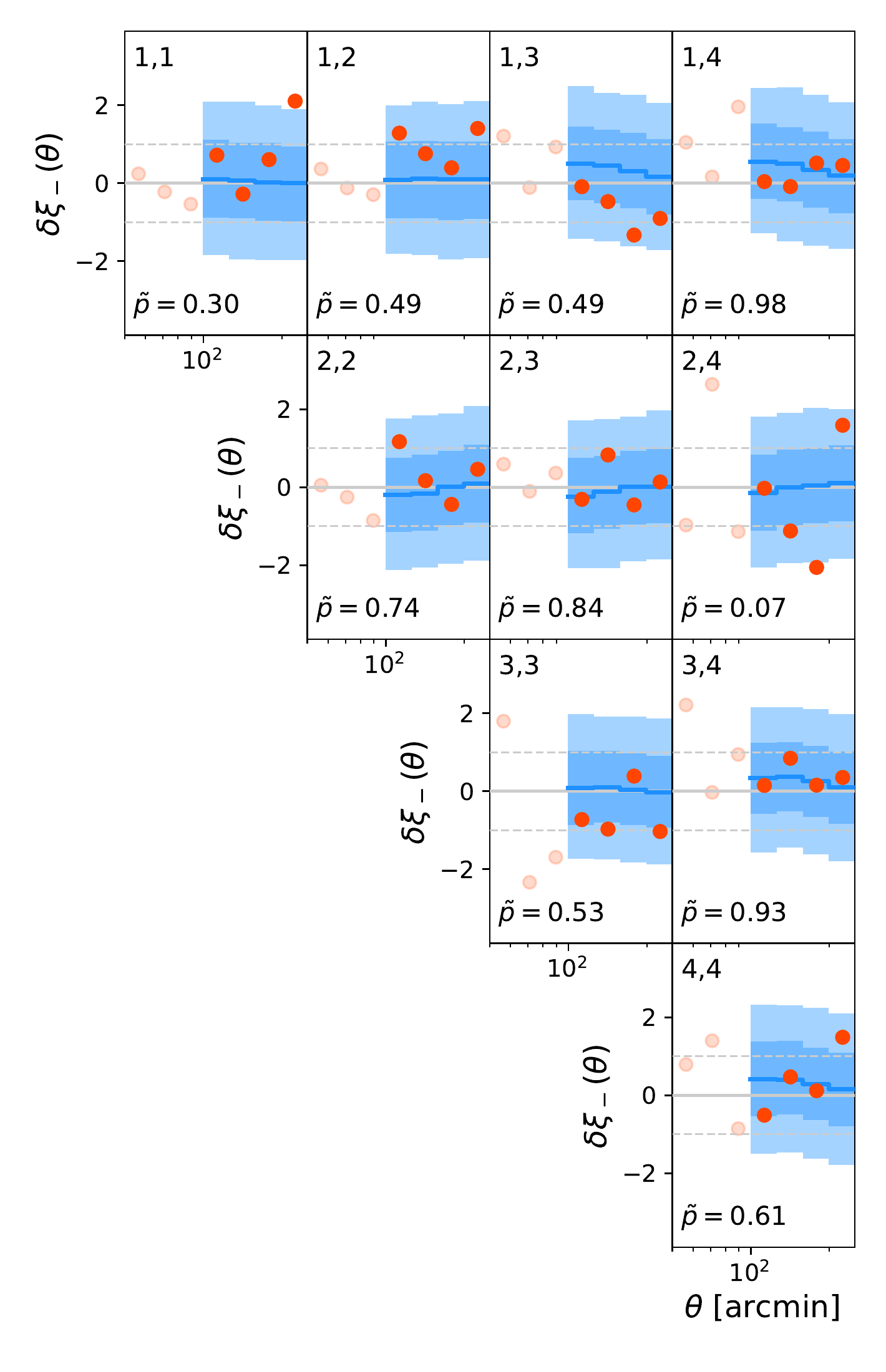}
    };
    \end{tikzpicture}
    \includegraphics[scale=0.4]{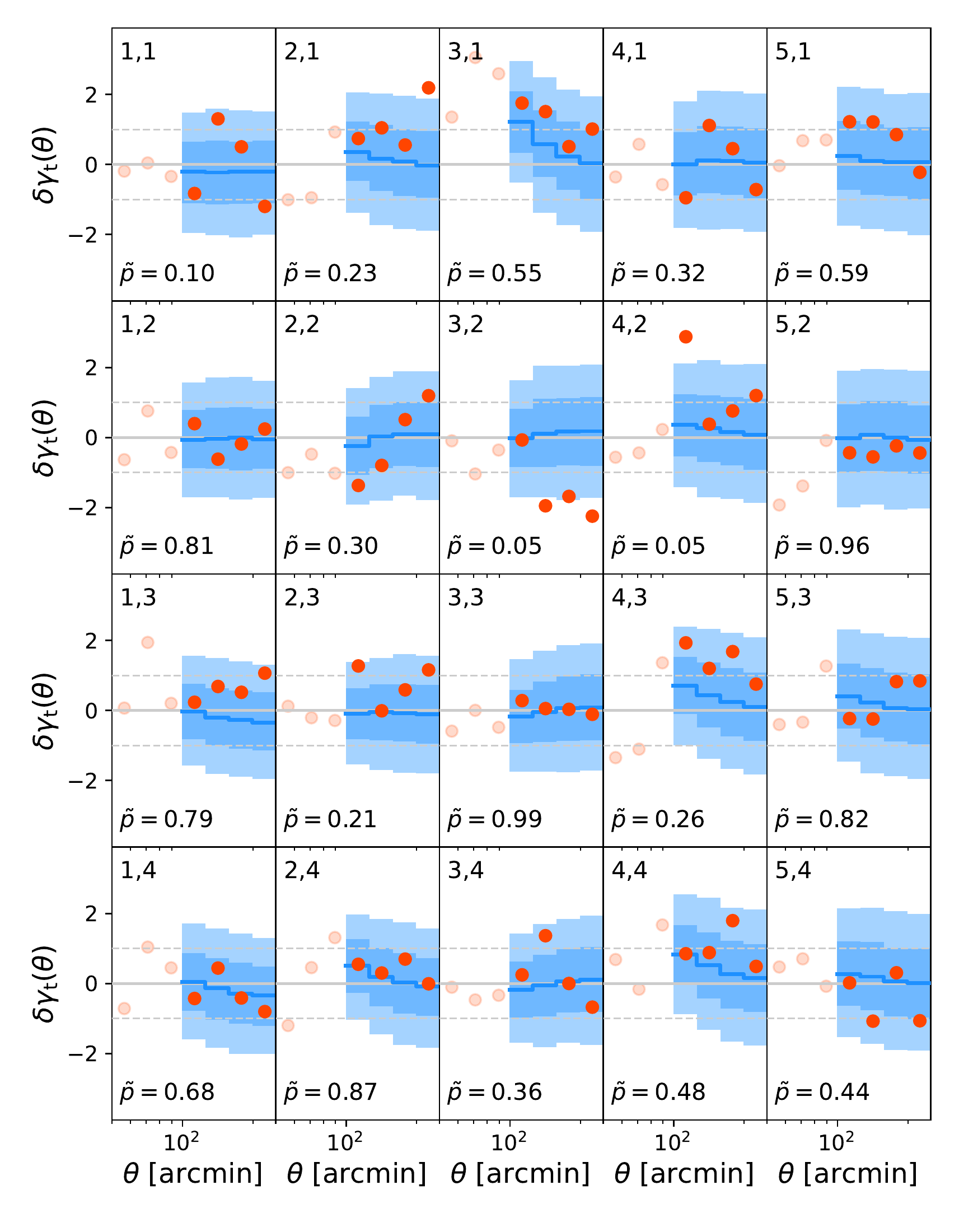}\\
    \includegraphics[scale=0.4]{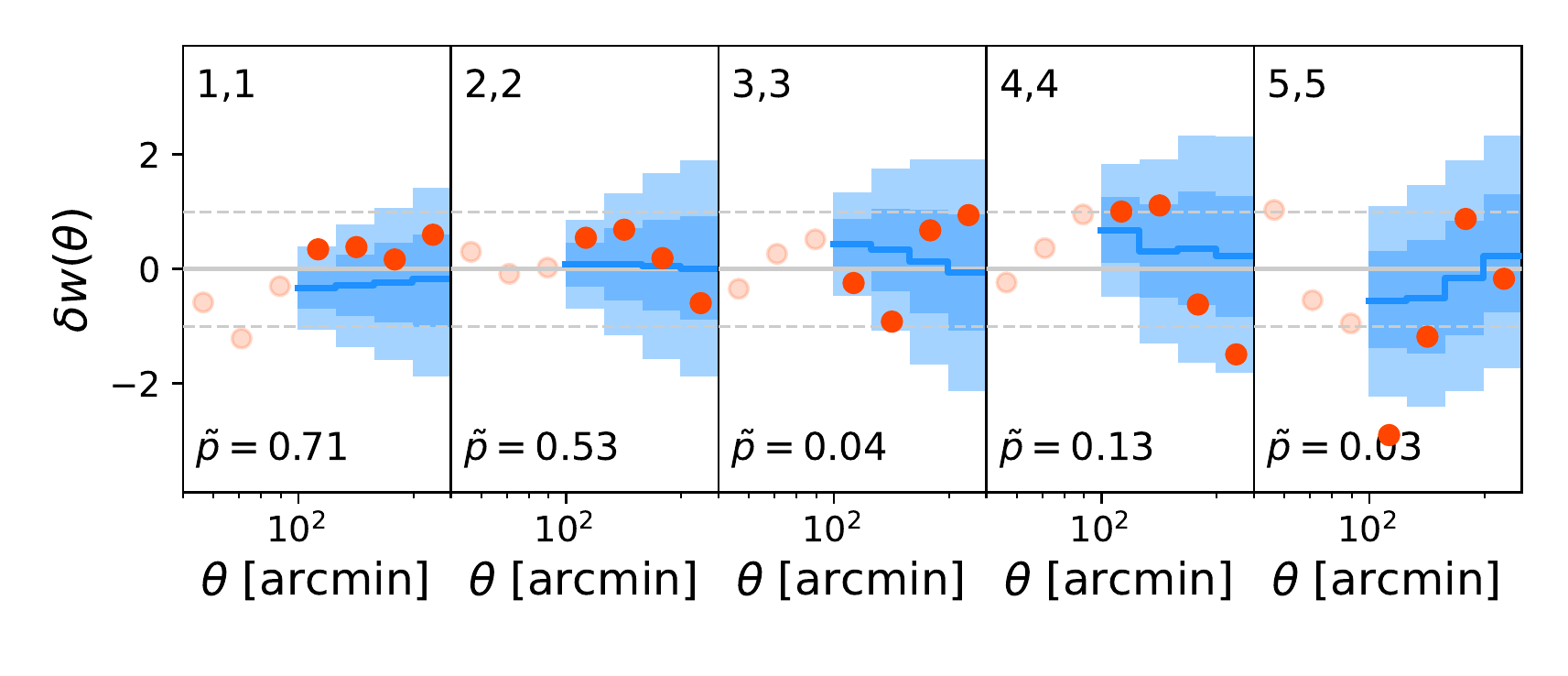}
    \end{center}
    \caption{PPD for all 3x2pt correlation functions at large scales ($\theta>\SI{100}{\arcminute}$) conditioned on observed correlation functions at small scales (\textit{i.e.} data points at separation angle $\theta$ within scale cuts and $\theta<\SI{100}{\arcminute}$).   See \cref{fig:ppd_3x2} for explanation of bands. \label{fig:large_vs_small}}
\end{figure*}

Like splits of the data on redshift bin, splits of the data on angular scale are motivated by both considerations of physics and systematic errors.  Departures of galaxy clustering from the assumed linear bias model, for instance, could lead to tension between the small and large-scale measurements.  The measurements of clustering at large scales are expected to be particularly susceptible to data systematics, such as dust extinction or varying observing conditions (fluctuations in depth, airmass, exposure time, width of the point spread function).  We therefore consider PPD realizations of the large-scale components of the \3x2pt data vector, conditioned on the small-scale measurements.  We choose $\theta=\SI{100}{\arcminute}$  as the splitting point, leaving four data points for each two-point function in the large-scale parts.

\Cref{fig:large_vs_small} shows the PPD realizations for the large-scale \3x2pt, conditioned on the small-scale measurements. We obtained an overall $\tilde{p}$-value of 0.034, relatively low compared to the full \3x2pt goodness of fit. Although some individual two-point functions show low $p$-values, only the (3,4) bin of $\xip$ shows a $p$-value below 0.01, and we do not observe any noticeable trend related to redshift based on $p$-values alone. When considering clustering alone (still conditioned on small-scale measurements of all observables), we obtain a $\tilde{p}$-value of 0.030. Individual bins of the clustering two-point functions show lower $p$-values at higher redshifts, which could point to systematic effects affecting large-scale clustering measurements, such as survey observing conditions. We note this is another example of consistency tests where the uncalibrated $p$-value is indeed biased low, as shown in \cref{fig:meta_pval}.

\subsection{Testing for cosmic shear systematics}
\label{sec:xim_xip}

We finally consider tension within the cosmic shear measurements by splitting the two components $\xip$ and $\xim$. This test is motivated by potential systematic effects as well as modelling considerations. On the systematics side, PSF leakage and shear-dependent selection biases can generate a B-mode pattern that affects each component differently ($\xip$ is related to the sum of E- and B-mode power spectra, while $\xim$ is related to their difference). On the modelling side, we note that $\xip$ and $\xim$ receive contributions from different physical modes at given angular separation $\theta$. In particular, $\xim$ receives contributions from smaller scales impacted by non-linear evolution and baryonic physics, which justify stricter scale cuts for $\xim$ than $\xip$. 
Since $\xip$ has a higher signal-to-noise ratio than $\xim$ for the scales we consider, we will apply the test to $\xim$ conditioned on the $\xip$ posterior.

\Cref{fig:xim_vs_xip} shows the PPD test in this case.  We measure a $\tilde{p}$-value of 0.186, indicating good agreement between both components of cosmic shear measurements.

\begin{figure}
    \centering
    \includegraphics[width=\columnwidth]{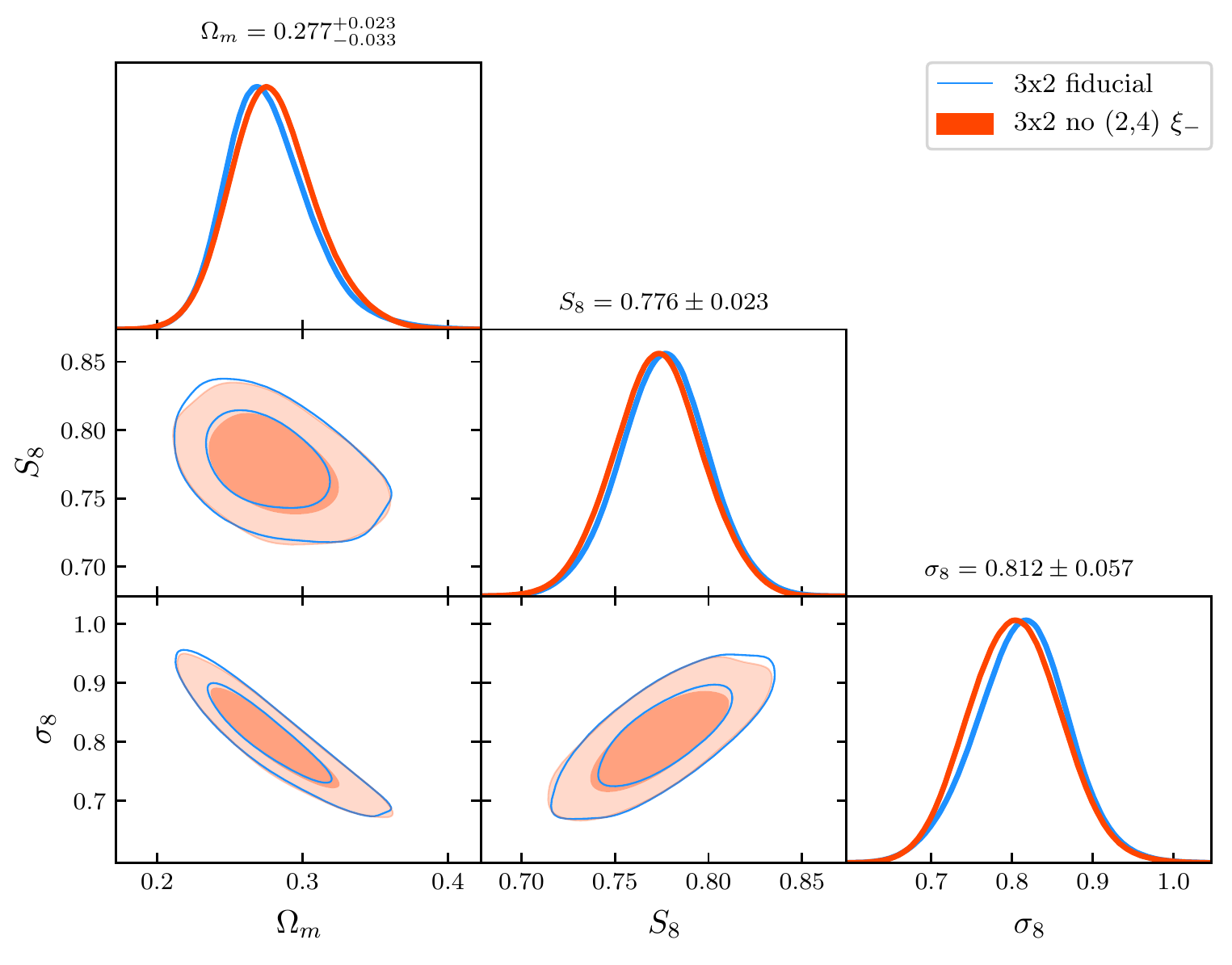}
    \caption{Impact on cosmological constraints of removing the $(2,4)$ $\xi_-$ bin from \3x2pt.}
    \label{fig:no_ximinus_2_4}
\end{figure}

\subsection{The impact of the $(2,4)$ bin of $\xi_{-}$}
\label{sec:bad_xim}

We consistently find in several tests described above that the $(2,4)$ redshift bin combination for $\xim$ yields a low $\tilde{p}$-value, even after calibration.  These tests are correlated (since they share parts of the same input data), so the fact that multiple tests show similarly low $\tilde{p}$-values is not surprising.  Moreover, as pointed out in \cref{sec:low_pvalues}, we have looked at many $\tilde{p}$-values for different redshift bin combinations, so the fact that one of them shows a low $\tilde{p}$-value is not unexpected.  Given that $\xim$ is sensitive to smaller scales and more susceptible to systematics than $\xip$, the fact that this bin shows the largest discrepancy between the observed data and that predicted by the PPD motivates us to verify its impact on the DES Y1 \3x2pt constraints.  When measuring the goodness of fit of the full \3x2pt vector, removing this bin increases the agreement between the model and the data, yielding a $\tilde{p}$-value of \num{0.13} ($p=0.10$ uncalibrated), compared to \num{0.065} for the full data vector.
\Cref{fig:no_ximinus_2_4} shows the impact on the DES Y1 \3x2pt cosmological constraints of removing the $(2,4)$ redshift bin combination of $\xim$.  We find that the impact on the cosmological constraints is negligible, essentially within the uncertainty of the sampling algorithm. We conclude that, while it may be the case that the $(2,4)$ bin combination of $\xim$ yields a bad fit to \lcdm, this measurement alone has little impact on the DES Y1 cosmological results.  We also note that this bin combination makes up only about 1.5\% of the total DES Y1 data vector.

\section{Internal consistency tests for DES Year 3}
\label{sec:y3}

In addition to testing the internal consistency of the DES Y1 joint probes analysis, one of the goals of this paper is to lay out the set of consistency tests that will be applied to the DES Y3 joint probes analysis. We plan on performing the same tests as those that were applied here to DES Y1 (summarized in \cref{tab:pvalues}), supplemented by additional probe-specific tests, such as testing large \vs small scales in cosmic shear alone. Importantly, we select the few most relevant tests to be performed on the unblinded data vector prior to any other comparison with our model. If these tests pass, we will allow comparing the data with the best-fit prediction and move forward to the parameter-level unblinding stage \citep[the blinding methodology is described in][]{2020MNRAS.494.4454M}. We keep the number of tests small in order to avoid redundancy and look-elsewhere effects (discussed in \cref{sec:pval_calib}). We therefore decide to fix a threshold at $p=0.01$ for each test used as part of the unblinding process and to select the following:
\begin{itemize}
    \item Goodness-of-fit tests: (1) full \3x2pt, (2) cosmic shear, (3) galaxy-galaxy lensing and clustering (referred to as 2x2pt);
    \item Consistency test: (4) cosmic shear \vs galaxy-galaxy lensing and clustering (2x2pt).
\end{itemize}
The three goodness-of-fit tests allow us to verify that the baseline model used in the analysis---including the cosmological model, but also models for intrinsic alignments, photometric redshifts, multiplicative bias and galaxy bias---is a good fit to the full data set as well as each individual probe. The consistency test allows us to verify that cosmic shear measurements are compatible with galaxy-galaxy lensing and clustering measurements (2x2pt) given the baseline model, indicating that it is sensible to combine them into the full \3x2pt analysis.
Because we \textit{a priori} do not expect Y3 data to be sensitive enough to rule out both $\Lambda$CDM and $w$CDM, we have chosen to allow revisiting the potential for bugs or flaws with non-cosmological parts of the modelling pipeline if we find the data incompatible even in $w$CDM, before seeing the final parameter values in any model.
Once the data have passed all unblinding criteria including those four tests, we will perform all other internal consistency tests, for which we will report calibrated $\tilde{p}$-values as well.

\section{Conclusion}
\label{sec:discussion}

In the context of mild to severe tensions between cosmological constraints on the \lcdm model reported by multiple experiments, it is crucial to assess the internal consistency of individual data sets.
In this paper, we have performed a series of internal consistency tests of the DES Y1 \3x2pt data using the posterior predictive distribution (PPD).  The PPD represents the distribution of possible (unobserved) data, conditioned on observed data, under a shared model.  The PPD tests have the advantage of performing comparisons directly in data space and are not impacted by prior volume effects, making it a particularly useful consistency test.  By comparing the PPD realizations to the true data, both with a $\chi^2$ test statistic and graphically, we assess the consistency of the DES data. We perform two kinds of tests: \textit{goodness-of-fit} tests to assess whether the model favored by the data is actually a good fit to DES Y1 \3x2pt measurements, and \textit{consistency} tests between disjoint subsets of the full data vector. In particular, we split the data vector into subvectors corresponding to different observables (cosmic shear, galaxy-galaxy lensing and clustering), measurements at small and large scales, and different redshift bins of cosmic shear data. The choice of $\chi^2$ test statistic yields conservative measures of consistency and we propose a calibration method to overcome exceedingly conservative $p$-values that may occur when the data splits result in too different posterior distributions. This method is applied consistently to all tests and we report such calibrated $\tilde{p}$-values throughout this analysis.

In general, we find that the DES Y1 \3x2pt data are self-consistent, and have an acceptable fit to \lcdm.  A direct graphical comparison of the PPD realizations to the true data yields no obvious discrepancies.  These results provides a strong validation of the DES Y1 measurements and cosmological constraints, as well as \lcdm.
However, there are a few peculiarities of the data. First, we find a somewhat low goodness-of-fit statistic for the full dataset of $\tilde{p}=0.065$ (${p = 0.046}$ uncalibrated). Secondly, we find a low $p$-value for the consistency test comparing large-scale to small-scale data elements (with a split at separation angle $\theta=\SI{100}{\arcminute}$) which suggests a small tension close to the $2\sigma$-level. This indicates either insufficient accuracy of the modeling of small-scale measurements or some observational systematic effect likely to impact large-scale measurements, potentially explaining the overall low $\tilde{p}$-value. Finally, we find that the $(2,4)$ bin combination of $\xim$ consistently yields a low $\tilde{p}$-value.  When this bin of $\xim$ is excluded, the $\tilde{p}$-value for the full \3x2pt data vector improves to $\tilde{p} = 0.13$.  However, excluding this bin from the analysis has negligible impact on the DES Y1 cosmological constraints.

The methodology developed here will be applied to the forthcoming analysis of the \3x2pt data vector measured from DES Y3 data.  The improvements in statistical noise with Y3 data make such tests even more interesting.
In particular, these tests will be essential to test the consistency with the cosmological model and look for unmodelled systematic effects. This would be even more relevant if the data were to show any sign of a real departure from the predictions of \lcdm.

\section*{Acknowledgements}

This paper has gone through internal review by the DES collaboration. The reviewers were Alex Alarcon, Andresa Compos and Youngsoo Park.

The authors would like to thank Masahiro Takada for fruitful discussions at early stages of the project, and Vivian Miranda and Scott Dodelson for useful comments and discussions. The authors would like to kindly thank the anonymous referee for their comments which helped us improve the paper.

Funding for the DES Projects has been provided by the U.S. Department of Energy, the U.S. National Science Foundation, the Ministry of Science and Education of Spain, 
the Science and Technology Facilities Council of the United Kingdom, the Higher Education Funding Council for England, the National Center for Supercomputing 
Applications at the University of Illinois at Urbana-Champaign, the Kavli Institute of Cosmological Physics at the University of Chicago, 
the Center for Cosmology and Astro-Particle Physics at the Ohio State University,
the Mitchell Institute for Fundamental Physics and Astronomy at Texas A\&M University, Financiadora de Estudos e Projetos, 
Funda{\c c}{\~a}o Carlos Chagas Filho de Amparo {\`a} Pesquisa do Estado do Rio de Janeiro, Conselho Nacional de Desenvolvimento Cient{\'i}fico e Tecnol{\'o}gico and 
the Minist{\'e}rio da Ci{\^e}ncia, Tecnologia e Inova{\c c}{\~a}o, the Deutsche Forschungsgemeinschaft and the Collaborating Institutions in the Dark Energy Survey. 

The Collaborating Institutions are Argonne National Laboratory, the University of California at Santa Cruz, the University of Cambridge, Centro de Investigaciones Energ{\'e}ticas, 
Medioambientales y Tecnol{\'o}gicas-Madrid, the University of Chicago, University College London, the DES-Brazil Consortium, the University of Edinburgh, 
the Eidgen{\"o}ssische Technische Hochschule (ETH) Z{\"u}rich, 
Fermi National Accelerator Laboratory, the University of Illinois at Urbana-Champaign, the Institut de Ci{\`e}ncies de l'Espai (IEEC/CSIC), 
the Institut de F{\'i}sica d'Altes Energies, Lawrence Berkeley National Laboratory, the Ludwig-Maximilians Universit{\"a}t M{\"u}nchen and the associated Excellence Cluster Universe, 
the University of Michigan, NFS's NOIRLab, the University of Nottingham, The Ohio State University, the University of Pennsylvania, the University of Portsmouth, 
SLAC National Accelerator Laboratory, Stanford University, the University of Sussex, Texas A\&M University, and the OzDES Membership Consortium.

Based in part on observations at Cerro Tololo Inter-American Observatory at NSF's NOIRLab (NOIRLab Prop. ID 2012B-0001; PI: J. Frieman), which is managed by the Association of Universities for Research in Astronomy (AURA) under a cooperative agreement with the National Science Foundation.

The DES data management system is supported by the National Science Foundation under Grant Numbers AST-1138766 and AST-1536171.
The DES participants from Spanish institutions are partially supported by MICINN under grants ESP2017-89838, PGC2018-094773, PGC2018-102021, SEV-2016-0588, SEV-2016-0597, and MDM-2015-0509, some of which include ERDF funds from the European Union. IFAE is partially funded by the CERCA program of the Generalitat de Catalunya.
Research leading to these results has received funding from the European Research
Council under the European Union's Seventh Framework Program (FP7/2007-2013) including ERC grant agreements 240672, 291329, and 306478.
We  acknowledge support from the Brazilian Instituto Nacional de Ci\^encia
e Tecnologia (INCT) do e-Universo (CNPq grant 465376/2014-2).

This manuscript has been authored by Fermi Research Alliance, LLC under Contract No. DE-AC02-07CH11359 with the U.S. Department of Energy, Office of Science, Office of High Energy Physics.

\section*{Data availability}

A general description of DES data releases is available on the survey website at \url{https://www.darkenergysurvey.org/the-des-project/data-access/}. DES Y1 cosmological data is available on the DES Data Management website hosted by the National Center for Supercomputing Applications at \url{https://des.ncsa.illinois.edu/releases/y1a1}. This includes the data vectors, redshift distributions and some of the posterior samples used in this analysis. The \texttt{CosmoSIS} software \citep{Zuntz:2015} is available at \url{https://bitbucket.org/joezuntz/cosmosis/wiki/Home}.

\bibliographystyle{mnras_2author}
\bibliography{refs} 




\appendix

\section{goodness-of-fit tests of individual observables}
\label{app:goodness_of_fit}

\begin{figure*}
\begin{center}
    \centering
    Cosmic shear goodness of fit \\
    \begin{tikzpicture}
    \node(a){\includegraphics[scale=0.4]{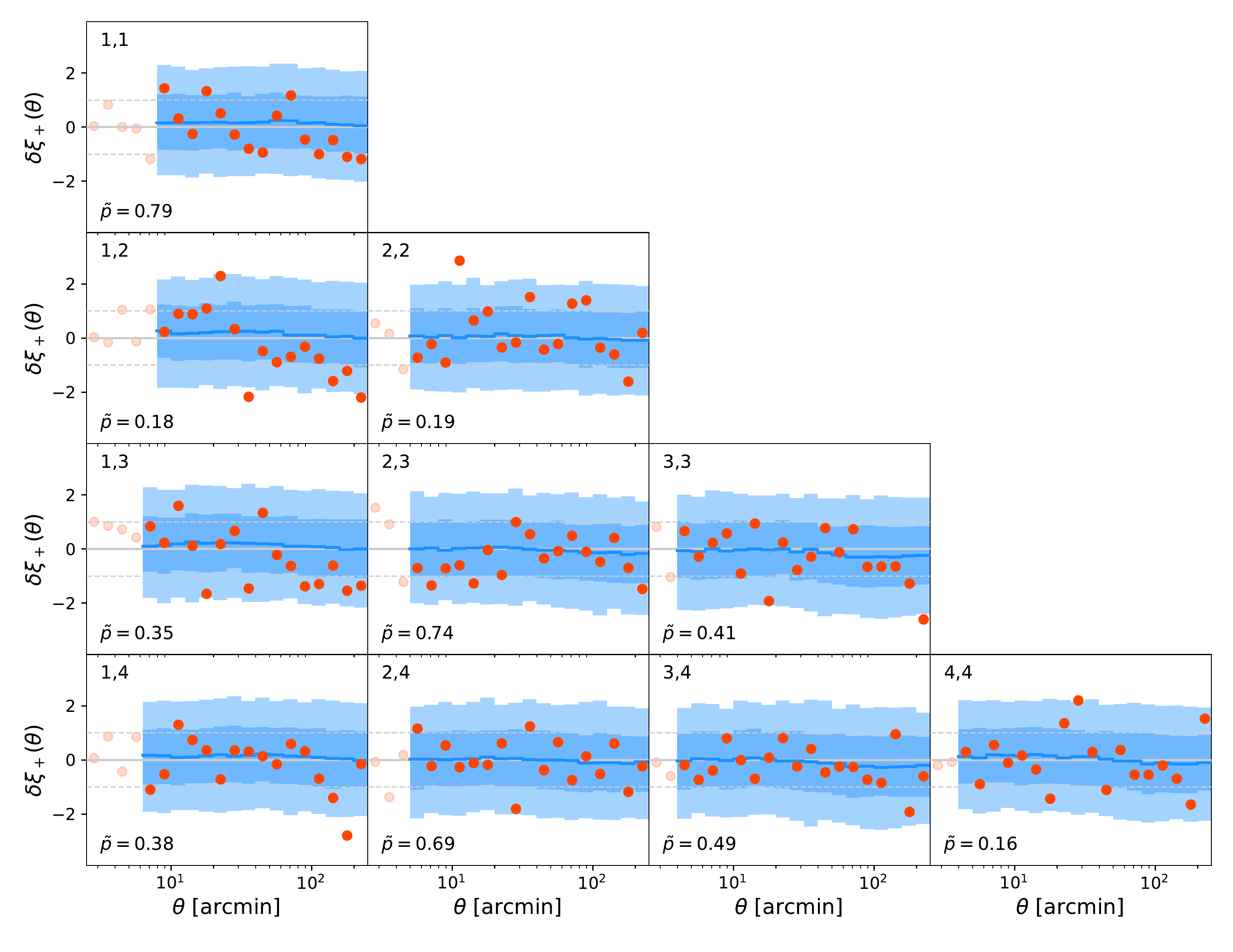}};
    \node at (a.north east)
    [
    anchor=center,
    xshift=-25mm,
    yshift=-35mm
    ]
    {
        \includegraphics[scale=0.4]{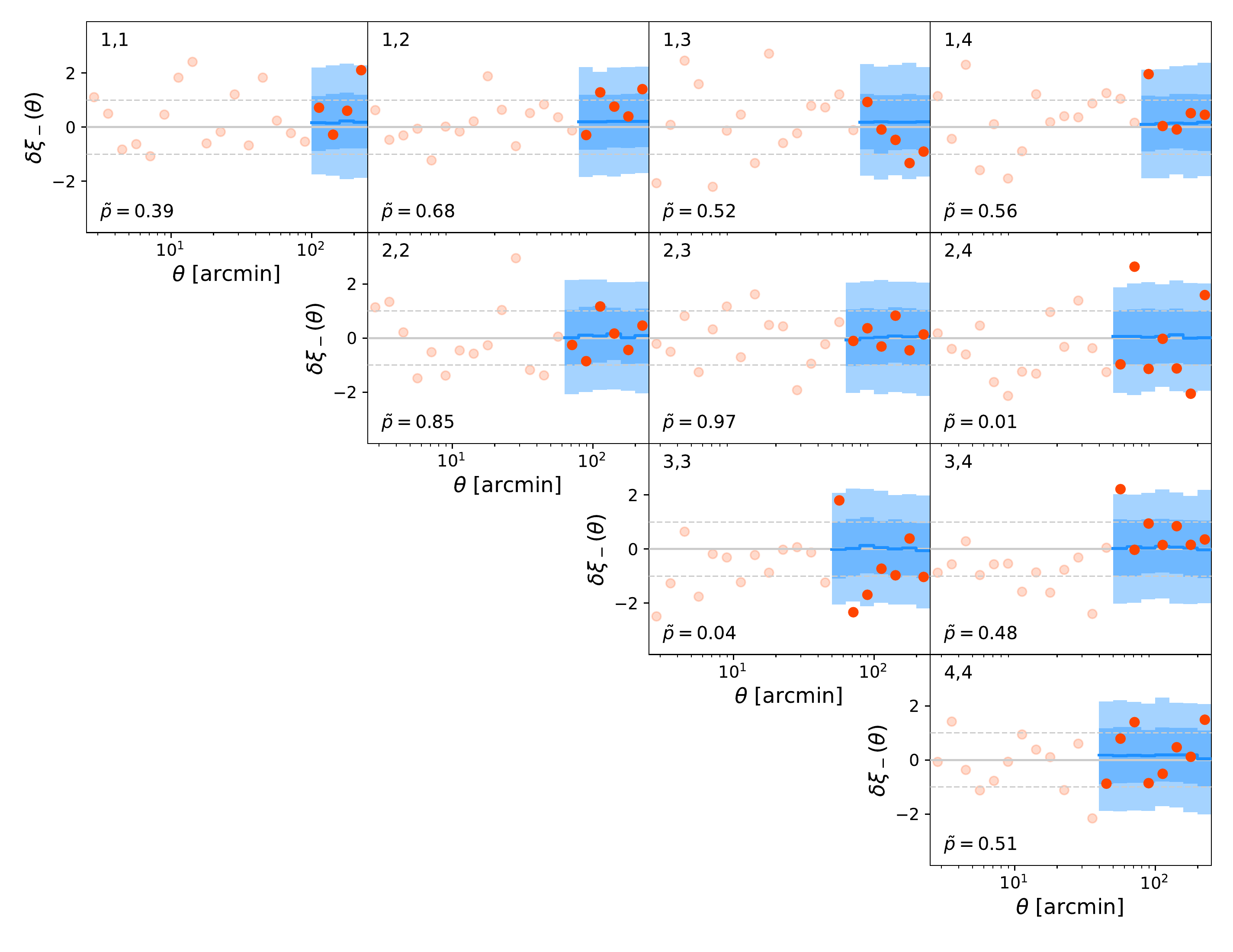}
    };
    \end{tikzpicture}
\end{center}
\caption{PPD goodness-of-fit test for cosmic shear alone.  See \cref{fig:ppd_3x2} for explanation of bands. \label{fig:gg_goodness_of_fit}}
\end{figure*}

\begin{figure*}
\begin{center}
Galaxy-galaxy lensing goodness of fit\\
\includegraphics[scale=0.4]{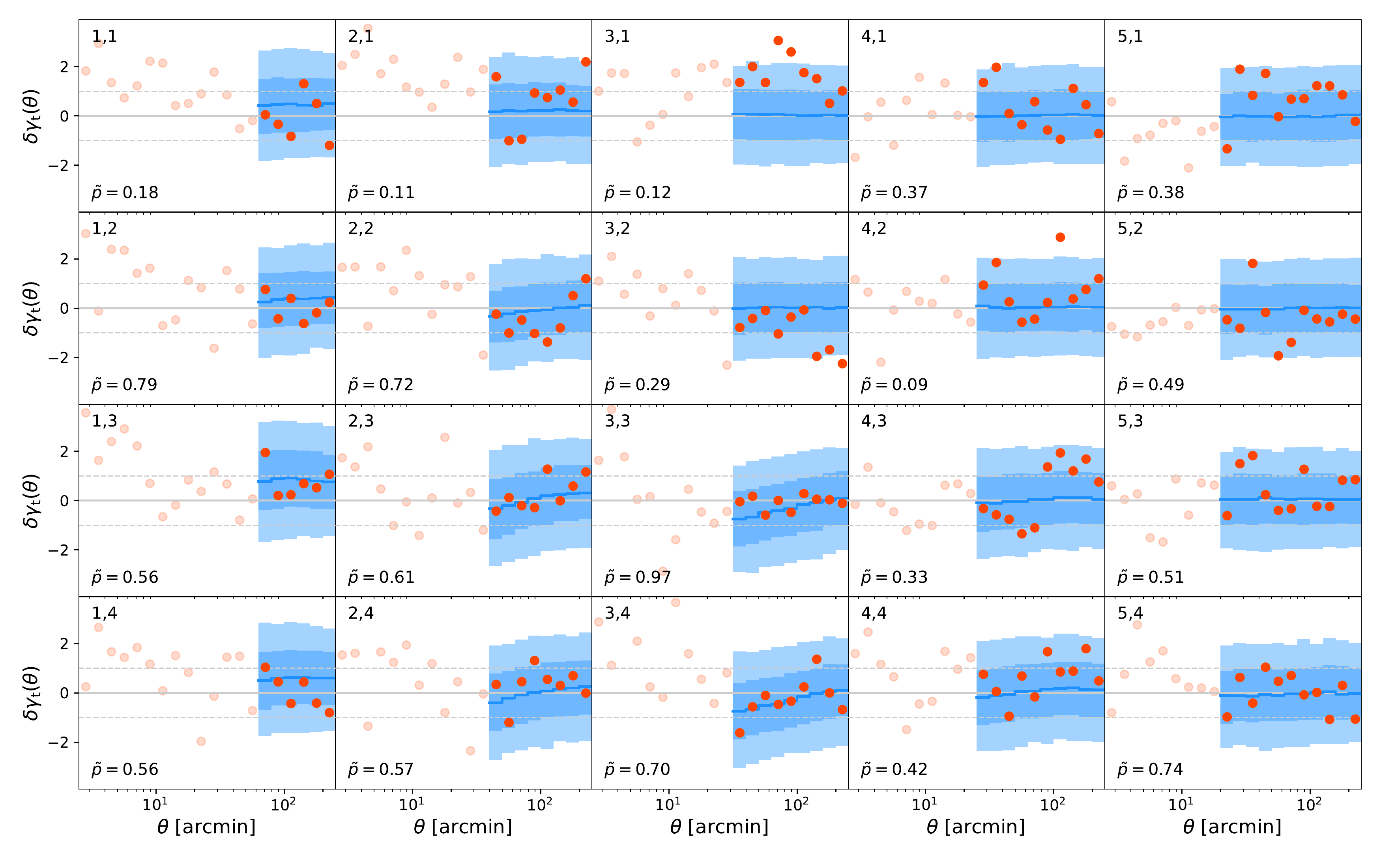}
\caption{PPD goodness-of-fit test for galaxy-galaxy lensing alone.  See \cref{fig:ppd_3x2} for explanation of bands. \label{fig:gammat_goodness_of_fit}}
\end{center}
\end{figure*}

\begin{figure*}
\begin{center}
Clustering goodness of fit\\
\includegraphics[scale=0.4]{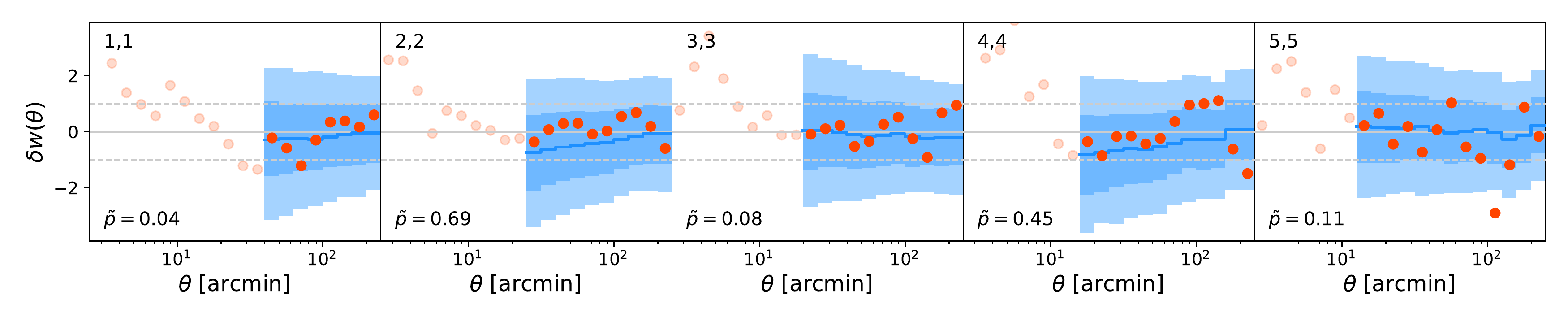}
\caption{PPD goodness-of-fit test for clustering alone.  See \cref{fig:ppd_3x2} for explanation of bands. \label{fig:wtheta_goodness_of_fit}}
\end{center}
\end{figure*}

In this section, we present PPD goodness-of-fit tests for individual DES Y1 \3x2pt probes---cosmic shear in \cref{fig:gg_goodness_of_fit}, galaxy-galaxy lensing in \cref{fig:gammat_goodness_of_fit} and clustering in \cref{fig:wtheta_goodness_of_fit}.  

\section{Additional conditional tests}
\label{app:conditioned}

\begin{figure*}
\begin{center}
Galaxy-galaxy lensing \vs cosmic shear and clustering\\
\includegraphics[scale=0.4]{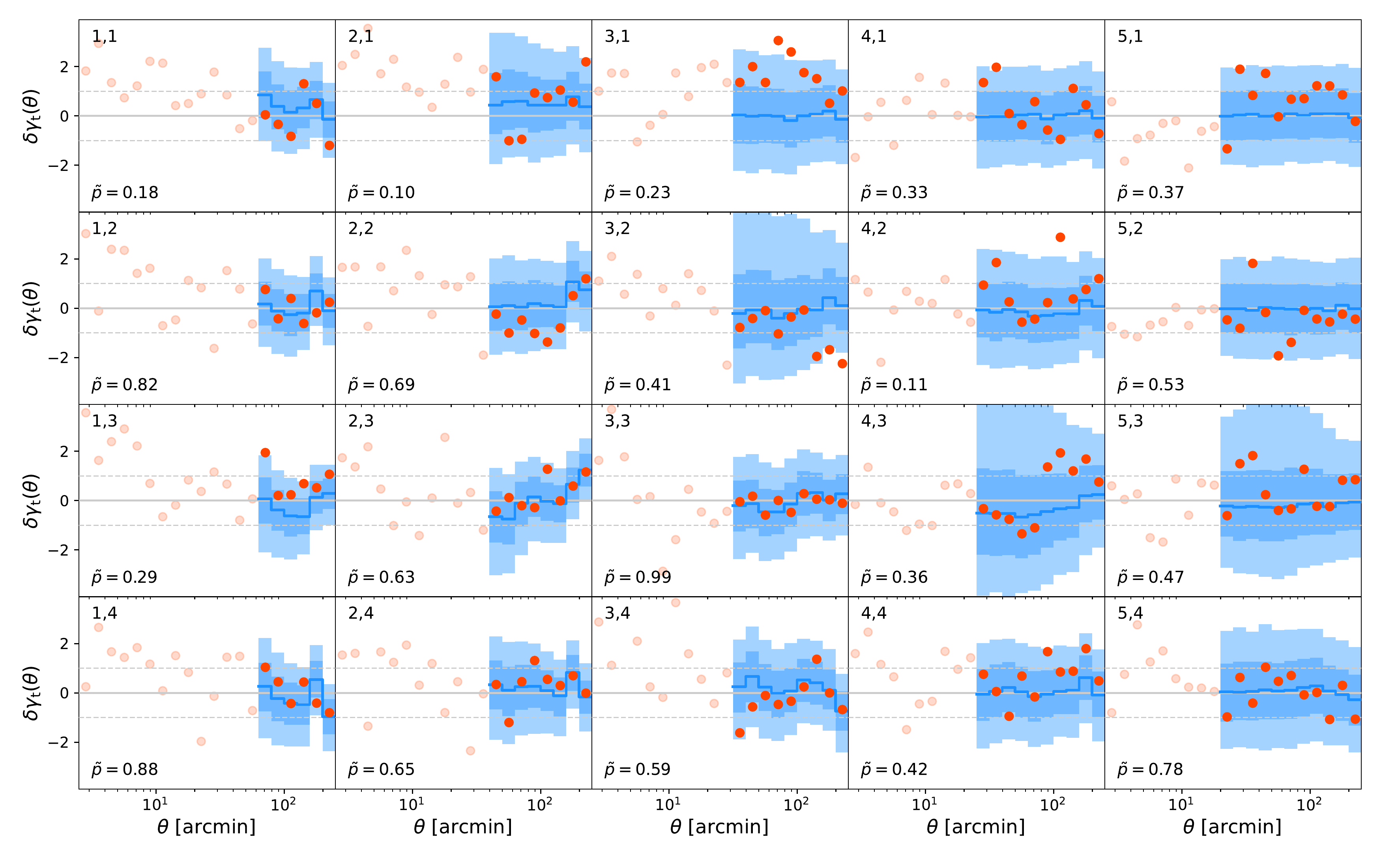}
\caption{PPD for galaxy-galaxy lensing, conditioned on the posterior from cosmic shear and clustering.  See \cref{fig:ppd_3x2} for explanation of bands. \label{fig:ppd_gammat_vs_ggwtheta}}
\end{center}
\end{figure*}

\begin{figure*}
\begin{center}
    \centering
    Cosmic shear \vs galaxy-galaxy lensing and clustering \\
    \begin{tikzpicture}
    \node(a){\includegraphics[scale=0.4]{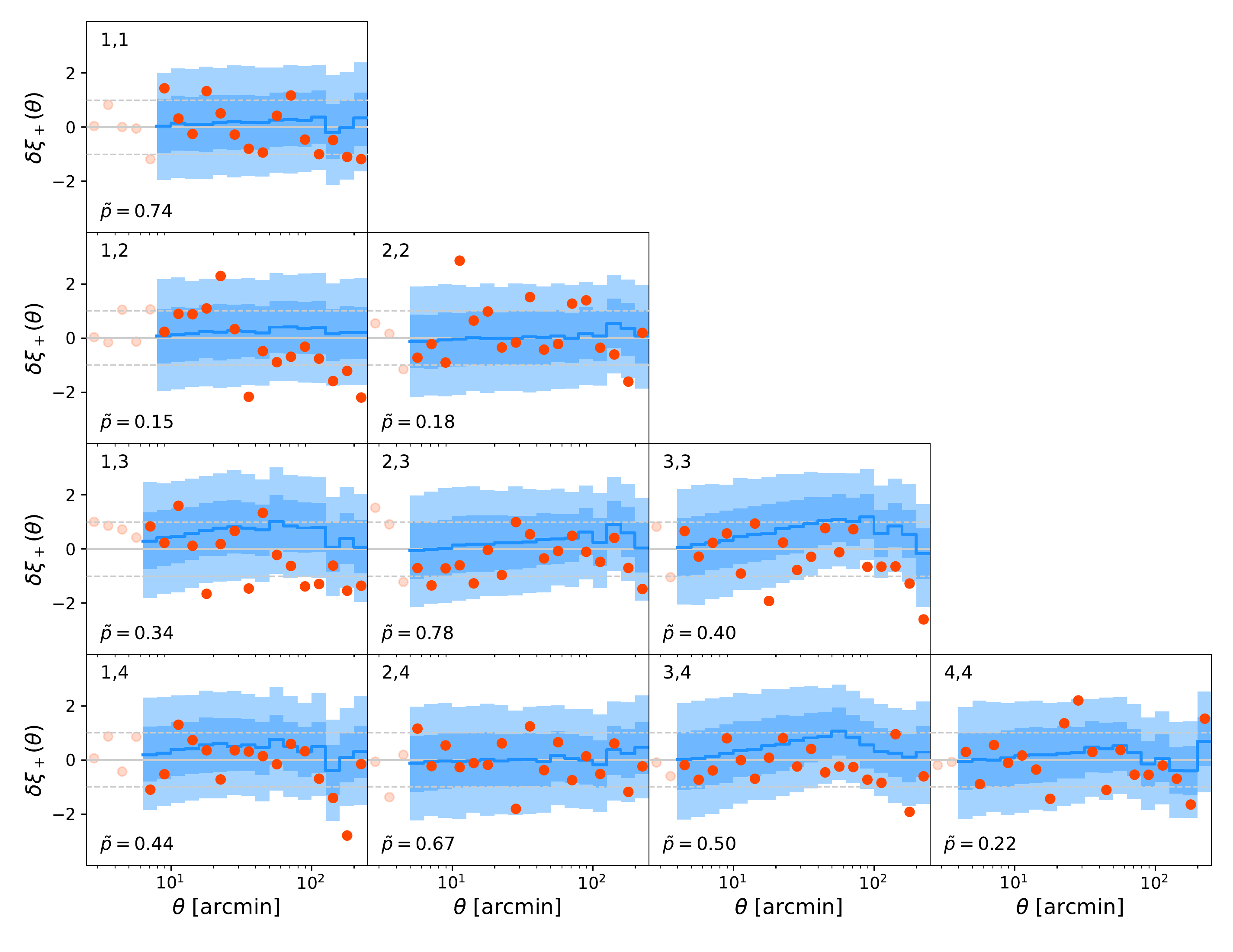}};
    \node at (a.north east)
    [
    anchor=center,
    xshift=-25mm,
    yshift=-35mm
    ]
    {
        \includegraphics[scale=0.4]{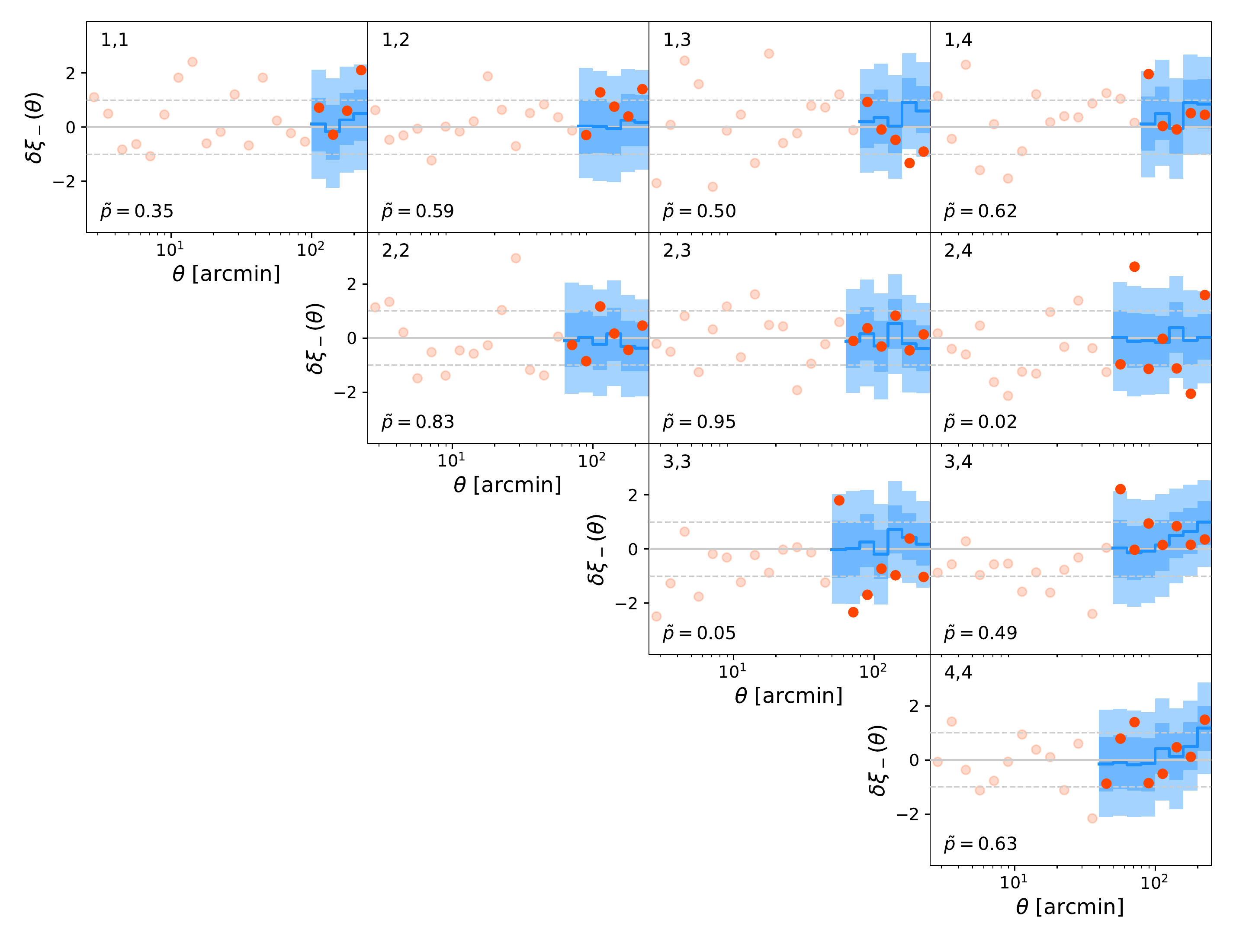}
    };
    \end{tikzpicture}
\end{center}
\caption{PPD for cosmic shear, conditioned on the posterior from galaxy-galaxy lensing and clustering.  See \cref{fig:ppd_3x2} for explanation of bands. \label{fig:ppd_gg_vs_wthetagammat}}
\end{figure*}

\begin{figure*}
\begin{center}
    \centering
    Cosmic shear $\xim$ \vs $\xip$\\
    \includegraphics[scale=0.4]{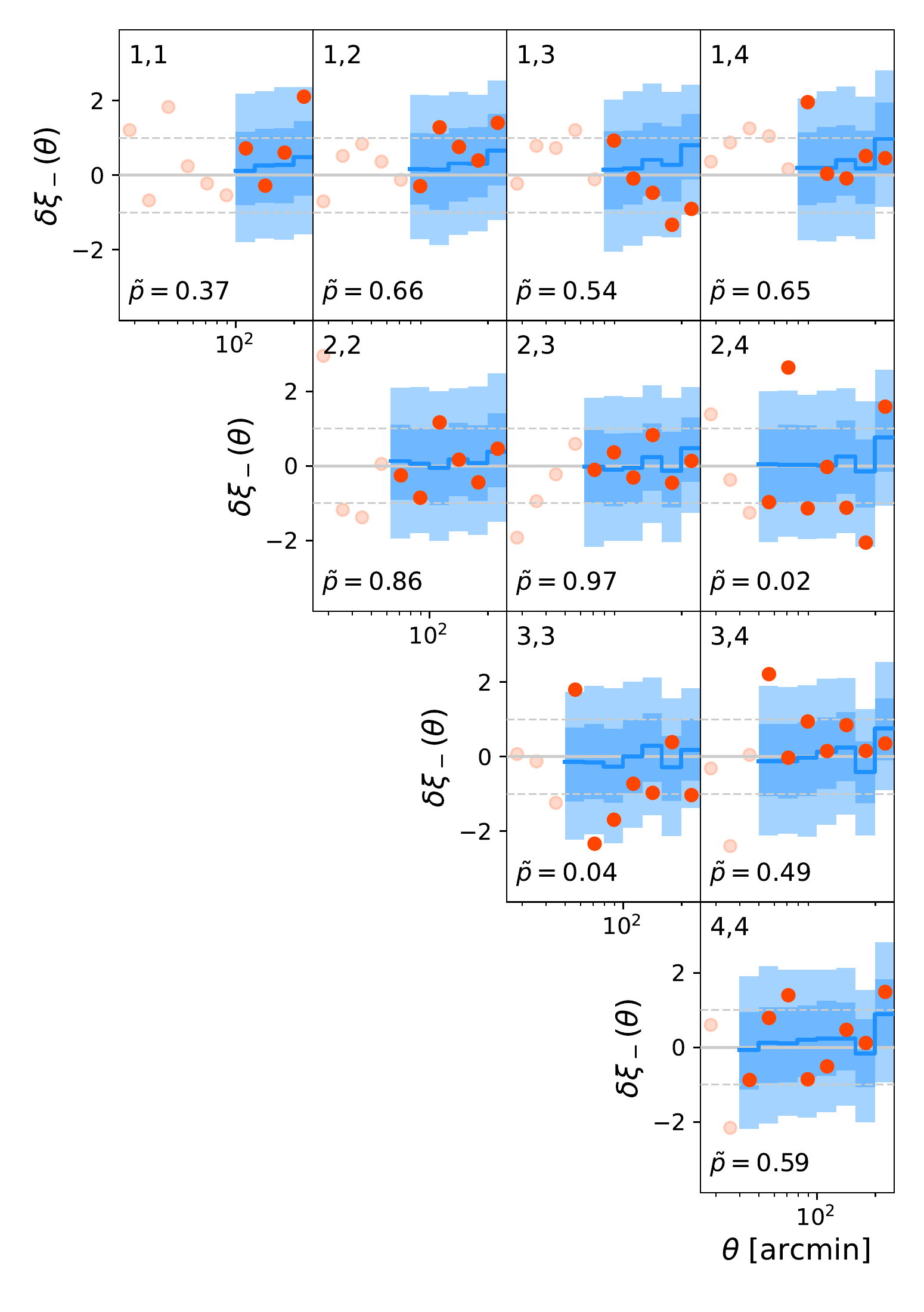}
\caption{PPD for $\xim$, conditioned on the posterior from $\xip$.  See \cref{fig:ppd_3x2} for explanation of bands. \label{fig:xim_vs_xip}}
\end{center}
\end{figure*}

In this section, we present additional results of PPD consistency tests. \Cref{fig:ppd_gammat_vs_ggwtheta,fig:ppd_gg_vs_wthetagammat} show, respectively, tests of galaxy-galaxy lensing and cosmic shear, conditioned on the two other probes. \Cref{fig:xim_vs_xip} shows the PPD test for cosmic shear $\xim$ conditioned on $\xip$ measurements.

\section{Calibration of PPD tests}
\label{app:meta_pval}

\begin{figure*}
    \centering
    \begin{minipage}[b]{0.3\textwidth}
    \centering
    3x2pt goodness of fit\\
    \includegraphics[scale=0.55]{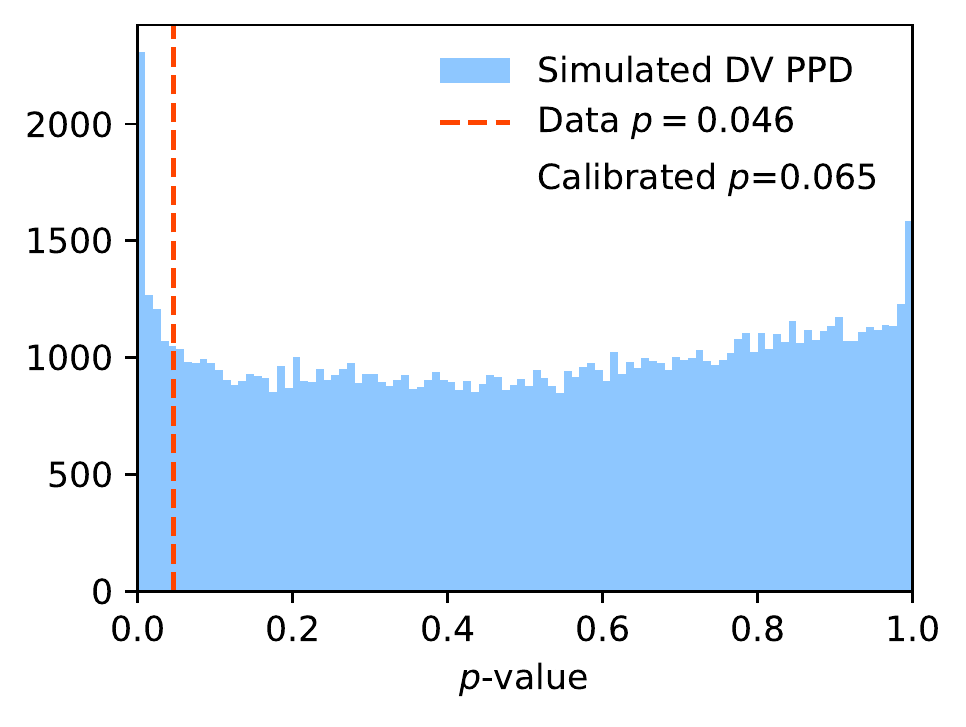}
    \end{minipage}
    \begin{minipage}[b]{0.3\textwidth}
    \centering
    3x2pt large \vs small scales\\
    \includegraphics[scale=0.55]{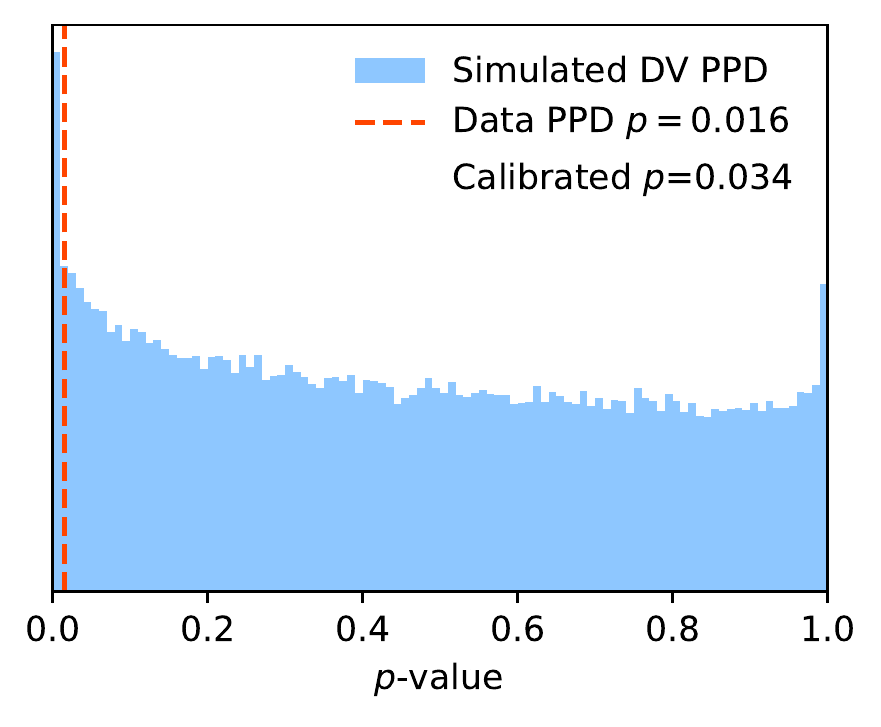}
    \end{minipage}
    \\
    \begin{minipage}[b]{0.3\textwidth}
    \centering
    Shear \vs galaxy-galaxy lensing+clustering \\
    \includegraphics[scale=0.55]{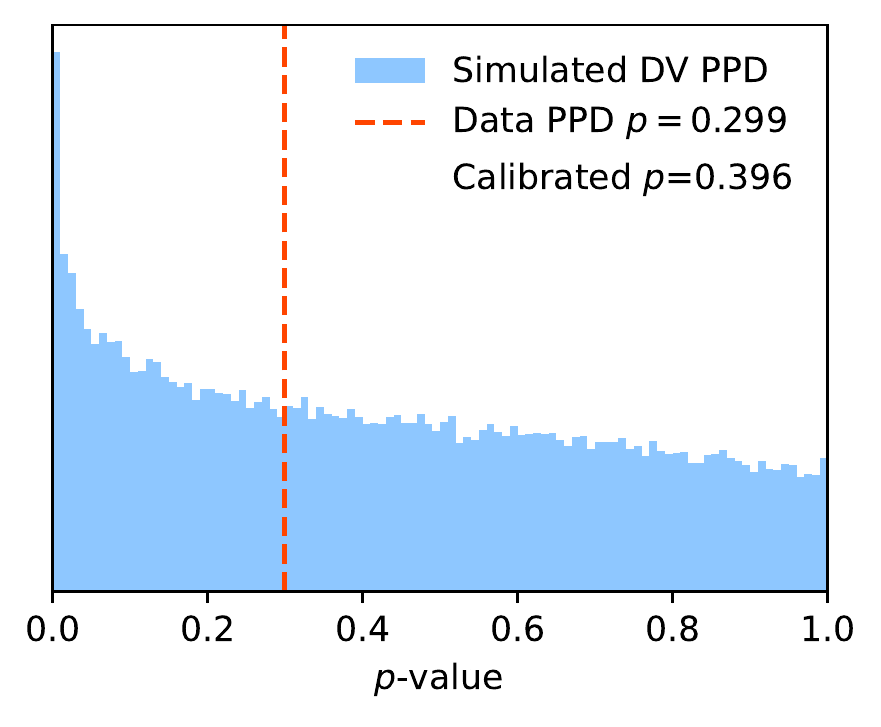}
    \end{minipage}
    \begin{minipage}[b]{0.3\textwidth}
    \centering
    Galaxy-galaxy lensing \vs shear+clustering \\
    \includegraphics[scale=0.55]{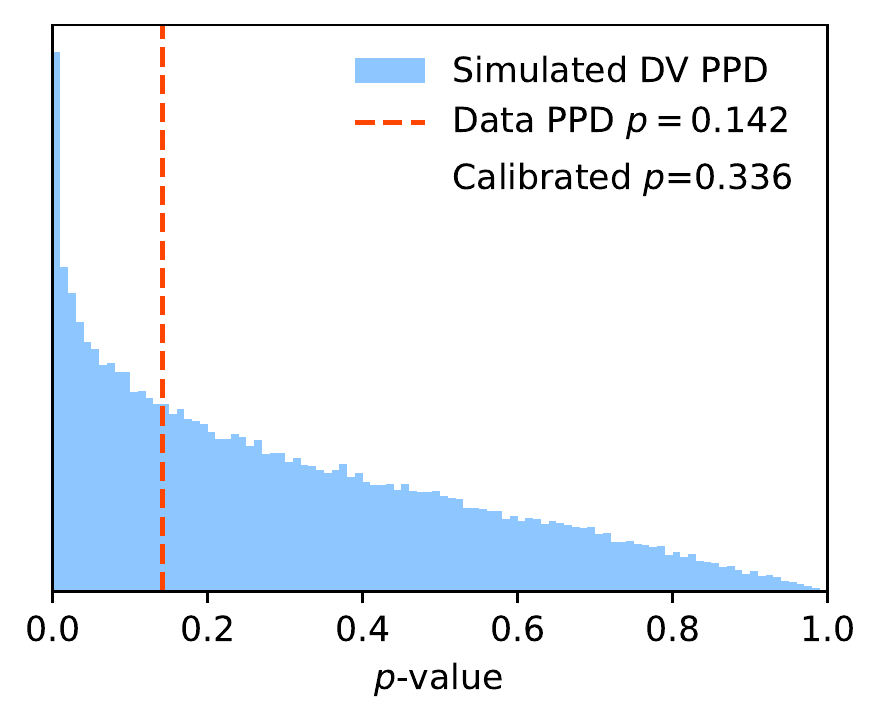}
    \end{minipage}
    \begin{minipage}[b]{0.3\textwidth}
    \centering
    Clustering \vs shear+galaxy-galaxy lensing\\
    \includegraphics[scale=0.55]{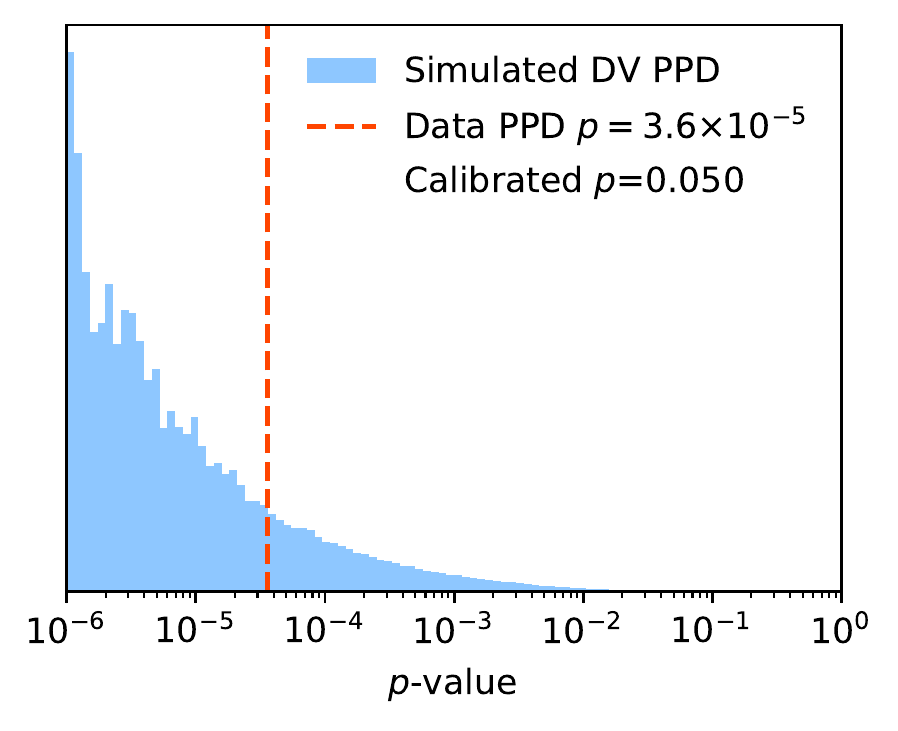}
    \end{minipage}
    \caption{Histograms of uncalibrated $p$-values (blue histograms) for simulated data vectors (DV) at the 3x2pt best-fit cosmology, compared to observed uncalibrated $p$-values from DES Y1 data (in red). Note the last panel uses a logarithmic scale as uncalibrated $p$-values are found to be very small, which is expected for the comparison of clustering \vs cosmic shear and galaxy-galaxy lensing.}
    \label{fig:meta_pval}
\end{figure*}

In this section, we present the distributions of uncalibrated $p$-values obtained for simulated data vectors generated at the 3x2pt best-fit cosmology, for five relevant PPD tests.
\Cref{fig:meta_pval} shows histograms these $p$-values---computed with importance sampling as explained \cref{sec:pval_calib}---for five cases: the 3x2pt goodness-of-fit test (\cref{sec:3x2_goodness_of_fit}), the 3x2pt large \vs small scales consistency test (\cref{sec:3x2_largescales_vs_3x2_smallscales}) and the three consistency tests between the two-point functions (\cref{sec:1x2pt_vs_2x2pt}). We compare these values to those obtained from the actual DES Y1 data (shown by the vertical red lines) and obtain \textit{calibrated} $\tilde{p}$-values given by the fraction of simulated $p$-values below the observed ones. As expected, the goodness-of-fit test presents a distribution very close to uniform, while those for consistency tests depart from uniformity, with a concentration of simulated $p$-values at low values that depends on the constraining power of the data splits on one another.

\section*{Affiliations}

$^{1}$ Department of Physics and Astronomy, University of Pennsylvania, Philadelphia, PA 19104, USA\\
$^{2}$ Institute for Astronomy, University of Hawai’i, 2680 Woodlawn Drive, Honolulu, HI 96822, USA\\
$^{3}$ Department of Physics \& Astronomy, University College London, Gower Street, London, WC1E 6BT, UK\\
$^{4}$ Department of Astronomy and Astrophysics, University of Chicago, Chicago, IL 60637, USA\\
$^{5}$ Kavli Institute for Cosmological Physics, University of Chicago, Chicago, IL 60637, USA\\
$^{6}$ Argonne National Laboratory, 9700 South Cass Avenue, Lemont, IL 60439, USA\\
$^{7}$ Kavli Institute for Particle Astrophysics \& Cosmology, P. O. Box 2450, Stanford University, Stanford, CA 94305, USA\\
$^{8}$ Department of Physics, Carnegie Mellon University, Pittsburgh, Pennsylvania 15312, USA\\
$^{9}$ Center for Cosmology and Astro-Particle Physics, The Ohio State University, Columbus, OH 43210, USA\\
$^{10}$ Institut de F\'{\i}sica d'Altes Energies (IFAE), The Barcelona Institute of Science and Technology, Campus UAB, 08193 Bellaterra (Barcelona) Spain\\
$^{11}$ Department of Physics, Stanford University, 382 Via Pueblo Mall, Stanford, CA 94305, USA\\
$^{12}$ SLAC National Accelerator Laboratory, Menlo Park, CA 94025, USA\\
$^{13}$ Institute of Astronomy, University of Cambridge, Madingley Road, Cambridge CB3 0HA, UK\\
$^{14}$ Kavli Institute for the Physics and Mathematics of the Universe (WPI), UTIAS, The University of Tokyo, Kashiwa, Chiba 277-8583, Japan\\
$^{15}$ Department of Astronomy, University of California, Berkeley,  501 Campbell Hall, Berkeley, CA 94720, USA\\
$^{16}$ Santa Cruz Institute for Particle Physics, Santa Cruz, CA 95064, USA\\
$^{17}$ D\'{e}partement de Physique Th\'{e}orique and Center for Astroparticle Physics, Universit\'{e} de Gen\`{e}ve, 24 quai Ernest Ansermet, CH-1211 Geneva, Switzerland\\
$^{18}$ Faculty of Physics, Ludwig-Maximilians-Universit\"at, Scheinerstr. 1, 81679 Munich, Germany\\
$^{19}$ Max Planck Institute for Extraterrestrial Physics, Giessenbachstrasse, 85748 Garching, Germany\\
$^{20}$ Universit\"ats-Sternwarte, Fakult\"at f\"ur Physik, Ludwig-Maximilians Universit\"at M\"unchen, Scheinerstr. 1, 81679 M\"unchen, Germany\\
$^{21}$ Department of Physics, Duke University Durham, NC 27708, USA\\
$^{22}$ Institute for Astronomy, University of Edinburgh, Edinburgh EH9 3HJ, UK\\
$^{23}$ Cerro Tololo Inter-American Observatory, NSF's National Optical-Infrared Astronomy Research Laboratory, Casilla 603, La Serena, Chile\\
$^{24}$ Departamento de F\'isica Matem\'atica, Instituto de F\'isica, Universidade de S\~ao Paulo, CP 66318, S\~ao Paulo, SP, 05314-970, Brazil\\
$^{25}$ Laborat\'orio Interinstitucional de e-Astronomia - LIneA, Rua Gal. Jos\'e Cristino 77, Rio de Janeiro, RJ - 20921-400, Brazil\\
$^{26}$ Fermi National Accelerator Laboratory, P. O. Box 500, Batavia, IL 60510, USA\\
$^{27}$ Instituto de Fisica Teorica UAM/CSIC, Universidad Autonoma de Madrid, 28049 Madrid, Spain\\
$^{28}$ Institute of Cosmology and Gravitation, University of Portsmouth, Portsmouth, PO1 3FX, UK\\
$^{29}$ CNRS, UMR 7095, Institut d'Astrophysique de Paris, F-75014, Paris, France\\
$^{30}$ Sorbonne Universit\'es, UPMC Univ Paris 06, UMR 7095, Institut d'Astrophysique de Paris, F-75014, Paris, France\\
$^{31}$ Department of Physics and Astronomy, Pevensey Building, University of Sussex, Brighton, BN1 9QH, UK\\
$^{32}$ Department of Astronomy, University of Illinois at Urbana-Champaign, 1002 W. Green Street, Urbana, IL 61801, USA\\
$^{33}$ National Center for Supercomputing Applications, 1205 West Clark St., Urbana, IL 61801, USA\\
$^{34}$ Physics Department, 2320 Chamberlin Hall, University of Wisconsin-Madison, 1150 University Avenue Madison, WI  53706-1390\\
$^{35}$ INAF-Osservatorio Astronomico di Trieste, via G. B. Tiepolo 11, I-34143 Trieste, Italy\\
$^{36}$ Institute for Fundamental Physics of the Universe, Via Beirut 2, 34014 Trieste, Italy\\
$^{37}$ Observat\'orio Nacional, Rua Gal. Jos\'e Cristino 77, Rio de Janeiro, RJ - 20921-400, Brazil\\
$^{38}$ Department of Physics, University of Michigan, Ann Arbor, MI 48109, USA\\
$^{39}$ Department of Physics, IIT Hyderabad, Kandi, Telangana 502285, India\\
$^{40}$ Institute of Theoretical Astrophysics, University of Oslo. P.O. Box 1029 Blindern, NO-0315 Oslo, Norway\\
$^{41}$ Institut d'Estudis Espacials de Catalunya (IEEC), 08034 Barcelona, Spain\\
$^{42}$ Institute of Space Sciences (ICE, CSIC),  Campus UAB, Carrer de Can Magrans, s/n,  08193 Barcelona, Spain\\
$^{43}$ Department of Astronomy, University of Michigan, Ann Arbor, MI 48109, USA\\
$^{44}$ Kavli Institute for Cosmology, University of Cambridge, Madingley Road, Cambridge CB3 0HA, UK\\
$^{45}$ School of Mathematics and Physics, University of Queensland,  Brisbane, QLD 4072, Australia\\
$^{46}$ Department of Physics, The Ohio State University, Columbus, OH 43210, USA\\
$^{47}$ Jet Propulsion Laboratory, California Institute of Technology, 4800 Oak Grove Dr., Pasadena, CA 91109, USA\\
$^{48}$ Center for Astrophysics $\vert$ Harvard \& Smithsonian, 60 Garden Street, Cambridge, MA 02138, USA\\
$^{49}$ Department of Astronomy/Steward Observatory, University of Arizona, 933 North Cherry Avenue, Tucson, AZ 85721-0065, USA\\
$^{50}$ Australian Astronomical Optics, Macquarie University, North Ryde, NSW 2113, Australia\\
$^{51}$ Lowell Observatory, 1400 Mars Hill Rd, Flagstaff, AZ 86001, USA\\
$^{52}$ Centre for Gravitational Astrophysics, College of Science, The Australian National University, ACT 2601, Australia\\
$^{53}$ The Research School of Astronomy and Astrophysics, Australian National University, ACT 2601, Australia\\
$^{54}$ Instituci\'o Catalana de Recerca i Estudis Avan\c{c}ats, E-08010 Barcelona, Spain\\
$^{55}$ Department of Astrophysical Sciences, Princeton University, Peyton Hall, Princeton, NJ 08544, USA\\
$^{56}$ Centro de Investigaciones Energ\'eticas, Medioambientales y Tecnol\'ogicas (CIEMAT), Madrid, Spain\\
$^{57}$ School of Physics and Astronomy, University of Southampton,  Southampton, SO17 1BJ, UK\\
$^{58}$ Computer Science and Mathematics Division, Oak Ridge National Laboratory, Oak Ridge, TN 37831\\


\bsp	
\label{lastpage}

\end{document}